\documentclass[preprint2]{aastex}
\usepackage{natbib}

\newcommand{\kms}{\ifmmode {\rm km\ s}^{-1} \else km s$^{-1}$\fi}
\newcommand{\Msun}{\ifmmode {\rm M}_{\odot} \else M$_{\odot}$\fi}
\newcommand{\Lsun}{\ifmmode {\rm L}_{\odot} \else L$_{\odot}$\fi}
\newcommand{\qo}{\ifmmode q_{\rm o} \else $q_{\rm o}$\fi}
\newcommand{\Ho}{\ifmmode H_{\rm o} \else $H_{\rm o}$\fi}
\newcommand{\ho}{\ifmmode h_{\rm o} \else $h_{\rm o}$\fi}
\newcommand{\ltsim}{\raisebox{-.5ex}{$\;\stackrel{<}{\sim}\;$}}

\newcommand{\vFWHM}{\ifmmode v_{\mbox{\tiny FWHM}} \else
                    $v_{\mbox{\tiny FWHM}}$\fi}
\newcommand{\CCF}{\ifmmode F_{\it CCF} \else $F_{\it CCF}$\fi}
\newcommand{\ACF}{\ifmmode F_{\it ACF} \else $F_{\it ACF}$\fi}
\newcommand{\Halpha}{\ifmmode {\rm H}\alpha \else H$\alpha$\fi}
\newcommand{\Hbeta}{\ifmmode {\rm H}\beta \else H$\beta$\fi}
\newcommand{\Hgamma}{\ifmmode {\rm H}\gamma \else H$\gamma$\fi}
\newcommand{\Hdelta}{\ifmmode {\rm H}\delta \else H$\delta$\fi}
\newcommand{\Lya}{\ifmmode {\rm Ly}\alpha \else Ly$\alpha$\fi}
\newcommand{\Lyb}{\ifmmode {\rm Ly}\beta \else Ly$\beta$\fi}
\newcommand{\HeI}{\ifmmode {\rm He}\,{\sc i}\,\lambda5876 \else 
	          He\,{\sc i}\,$\lambda5876$\fi}
\newcommand{\HeII}{\ifmmode {\rm He}\,{\sc ii}\,\lambda4686 \else 
	           He\,{\sc ii}\,$\lambda4686$\fi}

\newcommand{\hii}{H\,{\sc ii}}
\newcommand{\hei}{He\,{\sc i}}
\newcommand{\heii}{He\,{\sc ii}}

\newcommand{\feii}{Fe\,{\sc ii}}

\newcommand{\fevi}{Fe\,{\sc vi}}
\newcommand{\fevii}{Fe\,{\sc vii}}
\newcommand{\fex}{Fe\,{\sc x}}

\newcommand{\fexiv}{Fe\,{\sc xiv}}
\newcommand{\neiii}{Ne\,{\sc iii}}

\newcommand{\nev}{Ne\,{\sc v}}

\newcommand{\ciii}{\ifmmode {\rm C}\,{\sc iii} \else C\,{\sc iii}\fi}
\newcommand{\civ}{\ifmmode {\rm C}\,{\sc iv} \else C\,{\sc iv}\fi}
\newcommand{\Ni}{N\,{\sc i}}
\newcommand{\nii}{N\,{\sc ii}}

\newcommand{\oi}{O\,{\sc i}}
\newcommand{\oii}{O\,{\sc ii}}
\newcommand{\oiii}{O\,{\sc iii}}

\newcommand{\sii}{S\,{\sc ii}}
\newcommand{\siii}{S\,{\sc iii}}

\newcommand{\cav}{Ca\,{\sc v}}

\slugcomment{accepted for publishing in Astrophys.Journal}

\shorttitle{Narrow-Line Seyfert 1 Galaxy EUV Continuum}
\shortauthors{Dietrich et al.}

\begin{document}

%
%

\title{Probing the Ionizing Continuum of Narrow-Line Seyfert 1 Galaxies.
       I. Observational Results}

\author{
M.\,Dietrich
 \altaffilmark{1,2},
D.M.\,Crenshaw
 \altaffilmark{1},
and
S.B.\,Kraemer
 \altaffilmark{3}
}
\altaffiltext{1}
{Department of Physics and Astronomy, Georgia State University, 
 One Park Place South SE, Atlanta, GA 30303, USA.}
\altaffiltext{2}
{Department of Astronomy, The Ohio State University, 4055 McPherson Lab,
 Columbus, OH 43210, USA.}
\altaffiltext{3}
{Catholic University of America, NASA/Goddard 
 Space Flight Center, Code 681, Greenbelt, MD 20771, USA}
\email{dietrich@astronomy.ohio-state.edu}

\begin{abstract}
We present optical spectra and emission-line ratios of 12 Narrow-Line 
Seyfert\,1 (NLS1) galaxies that we observed to study the ionizing EUV 
continuum.
A common feature in the EUV continuum of active galactic nuclei is the big 
blue bump (BBB), generally associated with thermal accretion disk emission.
While Galactic absorption prevents direct access to the EUV range, it can be 
mapped by measuring the strength of a variety of forbidden optical emission 
lines that respond to different EUV continuum regions. 
We find that narrow emission-line ratios involving 
[\oii]\,$\lambda 3727$, H$\beta$, [\oiii]\,$\lambda 5007$, 
[\oi]\,$\lambda 6300$, H$\alpha$, [\nii]\,$\lambda 6583$, and 
[\sii]\,$\lambda 6716,6731$
indicate no significant difference between NLS1s and Broad-line Seyfert\,1
(BLS1) galaxies, which suggests that the spectral energy distributions of 
their ionizing EUV -- soft X-ray continua are similar.
The relative strength of important forbidden high ionization lines like 
[\nev]\,$\lambda 3426$ compared to \heii\,$\lambda 4686$ and the relative 
strength of [\fex]\,$\lambda 6374$ appear to show the same range as in BLS1 
galaxies. However, a trend of weaker 
F([\oi]\,$\lambda 6300$)/F(H$\alpha$) emission-line ratios is indicated for 
NLS1s compared to BLS1s.
To recover the broad emission-line profiles we used Gaussian
components. This approach indicates that the broad H$\beta$ profile can be 
well described with a broad component 
(${\rm FWHM} \simeq 3275 \pm 800$ km\,s$^{-1}$) 
and an intermediate broad component 
(${\rm FWHM} \simeq 1200 \pm 300$ km\,s$^{-1}$). The width of the broad
component is in the typical range of normal BLS1s. The emission-line flux that
is associated with the broad component in these NLS1s amounts to at least 
60\%\ of the total flux. Thus it dominates the total line flux, similar to 
BLS1 galaxies.
\end{abstract}
\section{Introduction}

It is generally supposed that a characteristic feature of active galactic 
nuclei (AGN), relating the X-ray and optical/UV regime, is the big blue bump 
(hereafter BBB) (e.g., Elvis et al.\,1986; Mathews \& Ferland 1987;
for a review see Collin 2001), which may be associated with thermal accretion 
disk emission (Rees 1984; Laor \& Netzer 1989; Band \& Malkan 1989; Sun \&
Malkan 1989; Laor et al.\,1997). 
The presence of the BBB, which is located in the EUV range, is indicated by 
upturns in the far-UV and soft X-rays. The strength and
location of the BBB likely depend on the black hole mass, the luminosity, 
and the accretion rate (e.g., Binette et al.\,1988,1989; Ross \& Fabian 1993;
Wandel 1999a,b).
However, because a direct observation of the EUV range is extremely difficult 
due to Galactic absorption, it is still an open question as to whether the BBB 
is as prominent as it is generally assumed or whether it is a shallower 
feature (Laor et al.\,1997).
The strength of a variety of emission lines in the optical domain and their
emission-line flux ratios that are sensitive to different parts of the ionizing
continuum in the EUV can be used as tracers of the continuum shape and strength
(Krolik \& Kallman 1988; Binette et al.\,1989; Netzer et al.\,1992; 
Zheng, Kriss \& Davidsen 1995; Korista, Ferland, \& Baldwin 1997; 
Alexander et al.\,1999; Kraemer \& Crenshaw 2000; Dietrich et al.\,2002).

Narrow-Line Seyfert\,1 galaxies (NLS1s) may hold key information about the 
structure and evolution of AGN. 
Although the permitted emission-line profiles in NLS1 spectra are unusually 
narrow (FWHM $\la 2000$ km\,s$^{-1}$), like in Seyfert\,2 galaxies, strong 
optical and ultraviolet \feii\ emission, distinct X-ray continuum properties, 
and rapid, large amplitude X-ray variations unmask them as type 1 Seyfert 
galaxies (Osterbrock \& Pogge 1985; Goodrich 1989; Boller et al.\,1997; 
Mathur 2000; Pogge 2000). 
Most of the observed properties of NLS1s can be explained in the framework of 
a low-mass black hole (few times $10^6 M_\odot$) and a high accretion rate 
that is close to the Eddington limit (e.g., Pounds, Done, \& Osborne 1996). 
This may explain the soft X-ray excess emission and rapid X-ray flux variations
(e.g., Boller et al.\,1996; Grupe et al.\,1995; Grupe et al.\,2000;
 Leighly 1999a).

NLS1 galaxies are prime candidates to investigate the strength and shape of the
soft X-ray -- EUV continuum, since they show generally a strong soft X-ray 
excess (Boller et al.\,1996; Leighly 2000) and also steeper hard X-ray continua
than normal Seyfert\,1 galaxies (Brandt et al.\,1997; Leighly 1999b). Recently,
Wang \& Netzer (2003) suggested that the soft X-ray properties of NLS1s are 
natural consequences of their super-Eddington accretion rates. However, only a 
few NLS1 galaxies (TON\,S180, Turner et al.\,2002; RE\,1034$+$39, 
Casebeer \& Leighly 2003), have been successfully observed in the EUV, 
revealing evidence for a BBB (although its strength is uncertain
due to uncertainties in the correction for Galactic extinction).

It has been suggested that broad-line Seyfert\,1 (BLS1) galaxies show a small 
or no BBB (Laor et al.\,1997; Zheng et al.\,1997; Telfer et al.\,2002). 
There are observational indications that the NLR of Seyfert\,2, Seyfert\,1, and
NLS1 galaxies might be different (e.g., Schmitt \& Kinney 1996; Schmitt 1998),
possibly caused by differences in the ionizing continuum SED.
However, to probe different EUV -- soft X-ray continua it is necessary to 
study emission lines that cover a wide range in ionization energies, at least 
up to $\chi _{ion} \simeq 100$\,eV 
(e.g., Alexander et al.\,1999; Kraemer \& Crenshaw 2000).
Studies using emission lines like [\oi], [\oii], [\oiii] reveal that different
ionizing continua for NLS1 and normal BLS1 galaxies have only a minor impact 
on these emission-line ratios (Nagao et al.\,2001). 
Thus, probing the ionizing continuum of NLS1s in the EUV employing emission 
lines that are sensitive to a wide energy range will provide information on 
the presence of a BBB and whether the BBB is stronger and shifted to higher 
energies compared to BLS1s, similar to the strong soft state of galactic black 
holes at high accretion rates (Pounds, Done, \& Osborne 1996;
Brocksopp et al.\,1999; Zdziarski et al.\,2002).

\section{Observations}

We have observed a sample of 12 bright nearby NLS1 galaxies using the 1.5-m 
telescope at Cerro Tololo Inter-American Observatory (Table 1). The 
observations have been obtained as part of the 
{\it Small and Moderate Aperture Research Telescope System} (SMARTS) 
collaboration. 
The NLS1 sample is restricted to bright AGN with $m_v \la 16^m$ to record
spectra with a signal-to-noise ratio of at least $S/N \simeq 30$ to 40 in the
continuum within a reasonable observing time.
We observed the NLS1 galaxies using the R-C spectrograph in longslit mode
on April 23 to April 27, 2003.
The observations were performed under non-photometric conditions except for the
last night when the spectra were taken under photometric conditions with
subarcsec seeing ($\sim 0\farcs 7$). For the first four nights the seeing
varied between $0\farcs 8$ and $2\farcs 0$, with $1\farcs 2$ on average which
had no effect due to the large slit width.

A Loral 1k CCD detector with $1200 \times 800$ pixels (pixel size 
$15\mu m \times 15 \mu m$) was used together with grating 26 (blue, 600 l/mm), 
grating 35 (red, 600 l/mm) together with blocking filter GG\,495, and grating 
36 (blue, 1200 l/mm) to observe spectra with higher spectral resolution in the 
H$\beta$ -- [\oiii]\,$\lambda \lambda 4959,5007$ range. In addition, these 
spectra of higher spectral resolution were valuable to intercalibrate the 
spectra observed for the blue and red optical wavelength range.
We used different settings for the blue grating to ensure that the redshifted
[\nev]\,$\lambda 3426$ emission line was measured for each of the NLS1s in our 
sample and also different settings for the red grating to record the 
H$\alpha +$[\nii]\,$\lambda \lambda 6548,6583$ emission line complex and the 
[\sii]\,$\lambda \lambda 6716,6731$ emission lines.
Together with the spectra of high spectral resolution for the H$\beta $ --
[\oiii]\,$\lambda \lambda 4959,5007$ region we covered the entire optical 
wavelength range $\sim 3400 - 7000$ \AA .
The slit orientation was east-west and the slit width was set to $4\farcs 0$
for accurate absolute flux calibration and to mitigate the effect of 
atmospheric refraction.
Due to the design of the 1.5-m R-C spectrograph the spectral resolution is 
nearly unchanged for slit widths in the range of $1\farcs 0$ to $4\farcs 0$.
In addition to the wide slit width the NLS1 galaxies were observed within less 
than two hours of the meridian to minimize the light loss caused by 
differential refraction.

The total exposure times for the NLS1 galaxies range from 30 to 60 minutes 
using the blue and red grating, split into two exposures. For the high spectral
resolution data two spectra with 20 minutes exposure time each were observed.
The standard stars EG\,274, LTT\,3218, LTT\,4816, and LTT\,7987 
(Hamuy et al.\,1992) were observed for relative flux calibration each night, 
with exposure times of 5 to 15 minutes in the blue and red wavelength range
and 15 to 20 minutes for the spectra with higher spectral resolution.

\section{Narrow-Line Seyfert 1 Spectra}
\subsection{Data Analysis}

The NLS1 galaxy and standard star spectra were processed using standard 
MIDAS\footnote{Munich Image Data Analysis System, trade-mark of the European
Southern Observatory} software. 
The night sky component of the 2\,D-spectra was subtracted by fitting third 
order Legendre polynomials perpendicular to the dispersion, along each spatial
row of the spectra using areas on both sides of the object spectrum which were
not contaminated by the NLS1 galaxies or other objects.
To extract 1\,D spectra we adjusted the extension of the spatial extraction 
windows to match the seeing recorded during the observation. We applied 
spatial extraction windows of sizes with $4\arcsec \times 17\arcsec$ to 
$4\arcsec \times 23\arcsec$ which includes a significant fraction of the
host galaxy spectra.
Helium-argon wavelength calibration spectra were taken at the beginning and the
end of each observing night.
The achieved spectral resolution amounts to 4.8\,\AA\ for both the blue and 
red wavelength range ($R\simeq 280$ km\,s$^{-1}$), with an internal error of 
the wavelength calibration of $\Delta \lambda \simeq 0.12$\,\AA . 
The spectra taken with grating 36 have a spectral resolution of 2.6\,\AA\
($R\simeq 160$ km\,s$^{-1}$).
The same spectral resolution for the employed gratings was derived examining
strong night sky emission lines ([\oi]\,$\lambda 5577$, 
[\oi]\,$\lambda 6300$, and [\oi]\,$\lambda 6364$). These night sky emission 
lines indicate an absolute uncertainty of the wavelength calibration of 
$\Delta \lambda \simeq 1.1$\,\AA\ ($\Delta {\rm v} \simeq 60$\,km\,s$^{-1}$).

We corrected each NLS1 spectrum for the atmospheric c-band and b-band 
absorption, caused by $O_2$ molecules using the standard star spectra that 
were observed during each night.
The spectra were corrected for atmospheric extinction applying the standard 
curve of La Silla (Schwarz \& Melnick 1993) and for Galactic extinction 
using the $E_{B-V}$ values of Dickey \& Lockman (1990) and the Galactic 
extinction curve of Savage \& Mathis (1979).
The sensitivity functions for the blue, red, and higher resolution spectra, 
based on individual standard stars, differ by less than $\sim 5$\,\%\ from the
mean sensitivity functions for the corresponding nights which we used to 
calibrate the NLS1 spectra.

The flux-calibrated NLS1 spectra have been transformed to the rest frame using
the corresponding redshift given in Table 1. We derived the redshift for each 
object by examining the emission-line profile of the strong 
[\oiii ]\,$\lambda \lambda 4959,5007$ emission
lines. A Gaussian profile was fit to the upper part of the line profile with at
least 50\,\%\ of the peak intensity. The rest-frame flux-calibrated spectra of 
the observed NLS1 galaxies are shown in Figure 1. 

\subsection{Multi-Component Fit Approach}

In contrast to luminous quasars, the host galaxy contribution in the optical, 
for less luminous Seyfert galaxies can be of the order of $\sim 50$\,\%\ in
the large aperture that we used. Hence, to measure emission line fluxes it is
important to correct for possible contamination, particularly to recover weaker
emission lines.
Almost all NLS1 galaxy spectra of our sample show significant host galaxy 
contributions to the observed continuum level as indicated by the presence of 
prominent stellar absorption features like CaII\,H and CaII\,K, the Mgb band, 
and the 
NaD absorption in the red wing of the \hei $\lambda 5876$ emission line.
Furthermore, the studied NLS1s show strong optical \feii\ emission features,
\feii $\lambda 4570$ and \feii $\lambda 5250$, blueward and redward of the 
H$\beta $ -- [\oiii]\,$\lambda\lambda 4959,5007$ complex (Fig.\,1).
Therefore, we analyzed each NLS1 spectrum employing a multi-component fit based
on  (i)   a power law continuum,
    (ii)  a template spectrum for the host galaxy contribution (Kinney et
          al.\,1996),
    (iii) a template for the \feii\ emission (Boroson \& Green 1992), and
    (iv)  a Balmer continuum emission spectrum (Grandi 1982; Storey \&
          Hummer 1995).

To construct representative host galaxy template spectra we combined several 
individual spectra of given Hubble types. These galaxy spectra were retrieved 
from the publicly available sample published by Kinney et al.\,(1996). The 
spectral resolution of the galaxy spectra amounts to $R\simeq 8$ \AA\ which is 
comparable to the spectral resolution of the NLS1 spectra that we observed.
The normal galaxy spectra were transformed into the rest frame. 
The atmospheric b-band absorption was corrected by linear interpolation. For 
elliptical galaxies we use templates based on NGC\,205, NGC\,2865, NGC\,1407, 
and a mean elliptical galaxy spectrum obtained by combining NGC\,221, 
NGC\,1399, NGC\,1404, NGC\,6868, and NGC\,7196.
The spectrum of NGC\,584 is used to represent a S0 type galaxy spectrum.
The Sa type galaxy template spectrum is given as an average of NGC\,1316 and
NGC\,6340, and the Sb type galaxy template spectrum is created using
NGC\,224 (M\,31) and NGC\,3031 (M\,81).

The I\,Zw1 \feii\ emission spectrum (Boroson \& Green 1992) is used as an 
\feii\ emission template. This \feii\ emission template was broadened in
accordance with the emission-line profile width of the broad permitted emission
line profiles. The width of the \feii\ emission lines was allowed to vary in
steps of $\Delta v = 250$ km\,s$^{-1}$ which is sufficient for a spectral
resolution of about $\Delta v \simeq 280$  km\,s$^{-1}$ that we achieved.
To compute Balmer continuum emission templates we followed the formalism that
was suggested by Grandi (1982) and for the higher order Balmer lines 
($n\leq 50$) we used the results presented by Storey \& Hummer (1995).

For the NLS1 galaxies of our sample we fit simultaneously a power law 
continuum, a host galaxy template, an \feii\ emission template, and a Balmer 
continuum template to the rest frame transformed spectra.
The Hubble type of the host galaxy was an additional free parameter. 
A first guess for the strength of the host galaxy was estimated based on 
stellar features like CaII\,H and CaII\,K, the CN g-band $\lambda 4300$\AA , 
the Mgb\,$\lambda 5175$ triplet, and the NaD absorption line that affects the 
red wing of the \hei $\lambda 5876$ emission-line profile.
The strength of the individual components of the multi-component fit are 
displayed in Figure 1 for each NLS1 under study.
In Table 2 we list the spectral slopes $\alpha$ of the power law continuum
fit ($F_\nu \propto \nu^\alpha$) together with the Hubble type of the host
galaxy as indicated by the best fit. 
While most of the NLS1 galaxies of this study show power law continuum slopes
in the optical that are in the typical range for AGN, i.e. $\alpha = - 0.2$ to
$- 1.4$ with $\alpha \simeq -0.8$ on average, four of these 12 NLS1 galaxies
show very steep power law continua ($\alpha \simeq 0.0$ to 1.9, Table 2). 
As can be seen in Figure 1 the host galaxy contribution has a significant 
impact on the slope of the observed continuum spectrum because these NLS1
galaxies were observed with a wide slit width that covered a significant part 
of the host galaxy.
To illustrate the host galaxy contribution to the observed NLS1 galaxy spectra 
we determined the relative flux at $\lambda = 5100$ \AA . 
The fraction of the host galaxy flux compared to the total observed continuum
strength at 5100 \AA\ is given in Table 2. It can be seen that the host galaxy
contributes at this wavelength generally $\sim 50$\,\%\ to the continuum flux 
for the NLS1 galaxies which we observed with a large aperture.

While the contribution of the host galaxy flux can be corrected by
employing a properly scaled galaxy template spectrum (Figure 1) there is an 
additional possibility of a contaminating starburst component. It has been 
pointed out that for numerous Seyfert\,2 galaxies star formation features are 
detected in their spectra (e.g., Cid-Fernandes et al.\,2001; 
Gon\c{c}alves et al.\,1999; V\'eron et al.\,1981a) while Seyfert\,1 galaxies 
are rarely accompanied by a starburst component (Rodriguez-Ardila \& Viegas 
2003). 
We can not rule out a presence of a young stellar population that would 
particularly affect the blue part of the spectrum, which might be expected 
if NLS1 galaxies are representing AGNs in an early evolutionary phase 
(e.g., Mathur 2000). While stellar features might be washed out by a strong 
emission-line spectrum, a steep power law continuum with spectral slopes in 
the range of $\alpha \simeq 0.25$ up to $\alpha \simeq 1.9$ 
(Mkn\,705, NGC\,4748, and RXS\,J20002-5417) can be mimicked by a blue 
continuum of a young stellar population (e.g., Leitherer et al.\,1999).
In addition to the multi-component fit analysis we scrutinized the 2-D 
spectra of the NLS1 galaxies for indications of extended line emission. While 
we find no indications for extended emission for Mkn\,734 and RXS\,J20002-5417,
most of the NLS1 galaxies show some very weak extended emission either for 
[\oiii]\,$\lambda \lambda 4959,5007$ and H$\beta$ or H$\alpha$ and 
[\nii]\,$\lambda 6583$. Only NGC\,4748 exhibits faint extended emission in the
blue and red part of the spectrum.
However, the intensity of the weak extended emission amounts to less than 
$\sim 1$\,\%\ to $\sim 1.5$\,\%\ compared to the peak intensity of the
corresponding emission lines. 
Furthermore, if H$\alpha$ and [\nii]\,$\lambda 6583$ can be detected their line
ratio is in the typical range for AGN excitation. Therefore, this weak 
additional line emission has no significant impact on the measured 
emission-line fluxes.

The final multi-component fit has been subtracted and the residuum spectra are 
plotted in Figure 1.
The NLS1 residuum spectra show that the correction for \feii\ emission and 
the underlying host galaxy spectrum uncovers particularly important emission 
lines like \heii\,$\lambda 4686$, as well as several [\fevi] and [\fevii] 
lines at $\lambda 5150 - 5200$ \AA .
Furthermore, the [\oiii]\,$\lambda \lambda 4959,5007$ emission-line profiles
which we use as profile templates for the NLR emission are corrected for
significant \feii\ contamination by the optical multiplet Fe42 (Phillips 1976).
The residuum spectra also illustrate that there is still significant \feii\ 
emission-line flux shortward of $\lambda \simeq 4250$ \AA\ (V\'eron-Cetty, 
Joly, \& V\'eron 2004), the blue end of the \feii\ emission template that we 
use for the spectral analysis.

\subsection{Separation of the NLR and BLR Flux Contribution}

To measure and analyze the flux of NLR emission lines, the BLR line emission 
contribution has to be removed. 
Generally, a broad emission-line profile can be described using a single 
or multiple Gaussian components (Clavel et al.\,1991; Koratkar \& Gaskell
1991; Kriss et al.\,1991) but also Lorentzian profiles are
suggested (see V\'eron et al.\,1980,1981a,b; Gon\c{c}alves et al.\,1999;
V\'eron-Cetty et al.\,2001). While for some AGNs, broad emission-line 
profiles can be well described with a single Lorentzian component, 
comparable and equally reasonable results can be achieved with Gaussian 
profiles, particularly with a combination of multiple components. 

To measure the narrow-line flux, a strong NLR emission line
can be used as a template profile. Using high-quality spectra of high 
spectral resolution, Whittle (1985a,b) has shown that NLR emission lines 
have similar profiles and especially that the Balmer emission-line 
profiles are identical with the [\oiii]\,$\lambda 5007$ line profile 
within the uncertainties, even in the case for strongly asymmetric 
profiles. Therefore, we have used the strong emission-line profiles of 
[\oiii]\,$\lambda \lambda 4959,5007$ to obtain a representative NLR line
profile template. 
To characterize the [\oiii]\,$\lambda \lambda 4959,5007$ profiles we 
have fit these lines using Gaussian profiles, as well as Lorentzian 
profiles. 
While the [\oiii] profiles can be well fit for each NLS1 galaxy under 
study using one or two Gaussian profiles, generally no reasonable fit
has been achieved employing Lorentzian profiles.
To use consistently the same profile type we fit both [\oiii] line 
profiles with generally two Gaussian components. In the case of
IRAS\,13224-3809 we use a single Gaussian component while for Mkn\,1239
the strong blue asymmetric [\oiii]\,$\lambda \lambda 4959,5007$ line 
profile requires a third Gaussian component.
In Figure 2 we display the reconstruction of the [\oiii]\,$\lambda 4959$ 
and [\oiii]\,$\lambda 5007$ emission-line profiles. These individual 
Gaussian components have a priori no physical meaning by themselves, but 
are only used to obtain a NLR line profile template.
To measure the narrow-line flux of important forbidden high ionization 
emission lines like [\nev]\,$\lambda 3426$, [\neiii]\,$\lambda \lambda 
3869,3968$, and [\fevii]\,$\lambda \lambda 5721,6087$, as well as to 
separate the BLR and NLR contributions of permitted emission lines,
we used the fitted [\oiii ] profile as the NLR emission-line profile 
template. 
While the NLR profile shape was kept fixed, the strength of the profile 
template was allowed to vary, as well as slight variations of the profile 
width and the location in wavelength space. This approach takes into
account that the profile width correlates with the excitation energy of 
the emission lines and that higher ionization lines tend to be blueshifted
relative to lower ionization lines (e.g., De\,Robertis \& Osterbrock 1984;
Whittle 1985b; Appenzeller \& \"{O}streicher 1988; Erkens et al.\,1997),
while the profile shape of the NLR emission lines is similar 
(Whittle 1985b).

To separate the NLR and BLR contributions of the broad permitted emission 
lines, we used the NLR template profile for the narrow-line emission and 
two Gaussian profiles to represent the broad emission-line profile. 
Following this approach we fit the broad H$\beta$ profile with an 
appropriately scaled NLR profile, and with a broad and an intermediate 
Gaussian component.
The deconvolution of the broad H$\beta$ emission-line profiles is 
displayed in Figure 2. 
It is evident that the line width of the intermediate component is 
generally significantly broader (on average about $2.8\pm0.3$ times) than
the narrow component which is associated with the NLR emission-line flux 
(Figure 2, Table 3).
However, for Mkn\,705, Mkn\,1239, and NGC\,4748 we detect a quite broad 
base of the [\oiii]\,$\lambda 5007$ emission line that contributes about 
$\sim 30$\,\%\ to the total [\oiii ]\,$\lambda 5007$ emission-line flux.
Although it is not necessarily the case that the 
[\oiii]\,$\lambda 5007$/H$\beta$ emission-line ratio is the same over the
range of the entire profile we used the NLR profile template based on
[\oiii]\,$\lambda 5007$ to recover the narrow H$\beta$ emission-line flux 
for Mkn\,705 and NGC\,4748. 
For these two NLS1 galaxies the width of the intermediate component 
of the broad H$\beta$ emission-line profile is still $\sim 25$\,\%\ 
broader than the broad base of the corresponding [\oiii] profile. In the 
case of Mkn\,1239 the strong blue asymmetry of the [\oiii]\,$\lambda 5007$ 
emission-line profile which is also detected by 
V\'eron-Cetty et al.\,(2001), is broader than the intermediate component 
of the H$\beta $ profile (Fig.\,2). 
Hence, for Mkn\,1239 we used only the narrow and intermediate component of
the [\oiii] profile fit to recover the narrow H$\beta$ emission-line flux.
For higher ionization lines of Mkn\,1239 like [\nev]\,$\lambda 3426$ this 
broad component of the [\oiii ] profile fit was fit independently.
In Figures 3 to 5 we show the results of the separation of the NLR and BLR
flux contributions for the H$\gamma +$\,[\oiii ]\,$\lambda 4363$, the 
\heii\,$\lambda 4686$, and H$\alpha +$\,[\nii ]\,$\lambda 6548,6583$ 
emission-line complexes.

\subsection{Gaussian versus Lorentzian Profile Components}

The choice of a specific profile type to fit the broad emission-line
profiles might have a significant impact on the measurement of the narrow 
emission-line flux.
Because we are using multiple Gaussian components we have investigated 
the range of uncertainty that might be introduced by our approach compared
to profile fits based, e.g. on a single Lorentzian profile, which could
potentially reduce the measured NLR flux 
(e.g., Gon\c{c}alves et al.\,1999; V\'eron-Cetty et al.\,2001). 
    
To estimate these systematic errors we analyzed the H$\beta$ emission
line profile of each NLS1 galaxy of this study employing a Lorentzian 
profile to fit the broad component. 
We studied two cases, assuming a narrow H$\beta$ emission-line flux 
strength as it is determined using a multiple Gaussian fit for the broad 
H$\beta$ component or assuming that we are overestimating the narrow 
H$\beta$ emission-line flux by a factor of 2. The accordingly scaled 
narrow-line H$\beta$ profile was subtracted from the entire H$\beta$ 
emission line. The resulting residuum was fit with a 
single Lorentzian profile. In Figure 6 the results of these tests are 
displayed for five representative NLS1 galaxies of our sample.
Although for a few NLS1 galaxies (IRAS\,13224-3809, Mkn\,1239, Mkn\,896,
RXS\,J20002-5417) a quite reasonable fit can be achieved, in general, we 
find that Lorentzian profiles yield less satisfying results than the 
approach we used, i.e., employing two Gaussian components (Figures 2 and 6).
Furthermore, the fit strongly depends on the wavelength range that is used
for a Lorentzian profile fit. If the fit is restricted to the inner part 
of the line profile, quite reasonable results can be obtained. However, 
the broad emission-line flux would be severely overestimated if these fits 
were extrapolated.  
If we include the H$\beta$ profiles wings into the fit range, generally 
the outer parts of the line core are underestimated while the core is 
significantly overestimated (Figure 6). 
Furthermore, using a single Lorentzian component, line profile asymmetries
are causing strong residuals that would require to employ  multiple 
Lorentzian components. 

Although the Lorentzian profile fits which are restricted to the inner
part of the profile are severely overestimating the flux of the profile 
wings we used these fits to estimate the uncertainties introduced by the 
specific choice of a profile type. 
The Lorentzian profile fit was subtracted from the original data. The
resulting residuum can be associated with the narrow-line H$\beta$
emission which is comparable to the narrow emission-line profile 
template (Figure 6). The estimated narrow H$\beta$ flux is on average
$\sim 10$\,\%\ higher than the flux we have determined, using a two 
Gaussian component fit. However, generally strong residua are left and 
for some NLS1s a broad base is present (Mkn\,705, Mkn\,896, NGC\,4748). 
For comparison, on the right side of Figure 6 we show the fit, which we 
obtained under the assumption that the narrow H$\beta$ emission-line flux 
might be up to $\sim 50$\,\%\ weaker than the flux we measured and hence 
result for some cases in rather low line ratios.
Although for some NLS1 galaxies of our sample a Lorentzian profile yields
a quite reasonable fit (IRAS\,13224-3809, Mkn\,1239, Mkn\,705, 
RXS,J20002-5417), the obtained results are rather poor for NLS1s showing 
profile asymmetries (ESO\,399IG20, IRAS\,15091-2107, Mkn\,291, Mkn\,783, 
NGC\,4748). 

Thus, the approach of using two Gaussian profiles
to fit the broad emission-line profile is more than reasonable instead of
using a mix of different profile types and that the introduced systematic 
uncertainty is of the order of $\sim 10$\,\% . It also should be noted
that the broad-line profiles show clear inflections in their profile
shape which cannot be described with a single component, indicating the 
presence of two components, independent from the type of profile used to fit.

\section{Results}

\subsection{Broad Emission-Line Profiles}

The broad H$\beta$ emission-line profile fit was used as a template to recover
the broad emission-line flux of H$\alpha$ and higher order Balmer lines.
We used the broad H$\beta$ emission-line profile fit as a first guess to 
recover the broad \hei $\lambda 5876$ and \heii\,$\lambda 4686$ line profiles
with a two Gaussian component fit. 
In Table 3 we present the results of the analysis of the broad H$\beta $ and 
\heii\,$\lambda 4686$ emission-line profiles.
We find that the NLS1 galaxies of our sample show a strong broad profile base 
for the permitted lines (Figure 2 and Table 3).
Although the deconvolution into two Gaussian components should be taken
cautiously, the width of the broad component indicates that NLS1 galaxies
contain gas at velocities which are in the typical range of BLS1 galaxies.
Furthermore, there is also the trend that the \heii\,$\lambda 4686$ emission 
line shows a significantly broader line profile than the hydrogen Balmer lines 
(Table 3). On average the broad component of the \heii\,$\lambda 4686$ 
emission line is about $\sim 40$\,\%\ wider than the broad H$\beta $ component 
($4720 \pm 460$ km\,s$^{-1}$ and $3275\pm 800$ km\,s$^{-1}$, respectively). 
These NLS1 galaxies show a similar trend of broader \heii\ emission like 
normal broad-line Seyfert\,1 galaxies (BLS1s) (e.g., Osterbrock 1993). 
While the broad components of the permitted emission-line profiles of H$\beta $
and \heii\,$\lambda 4686$ indicate an ionization stratification of the BLR, 
the 
intermediate broad component of these line profiles both show consistently on 
average a width of ${\rm FWHM} \simeq 1100$ to 1250 km\,s$^{-1}$ (Table 3).
The measured width of the intermediate broad profile amounts to 
${\rm FWHM}({\rm H}\beta _{im}) = 1125 \pm 335$ km\,s$^{-1}$ and
FWHM(\heii $ _{im}$) = $1245 \pm 330$ km\,s$^{-1}$.
These intermediate permitted line components display typical NLS1
profile widths. It should be noted that with the exception of CTS\,J13.12, the 
NLS1 galaxies of this study show wider broad components than the classical 
definition of a NLS1 galaxy (Osterbrock \& Pogge 1985) which is met by the 
width of the intermediate component (Table 3).
The presence of high velocity gas in the BLR of these NLS1 galaxies is further
supported by the full width at zero intensity (FWZI) of the H$\beta $ emission
line profile.
Since the FWZI of a broad emission-line profile depends crucially on the 
constraints given by the assumed continuum level, we only provide 
estimates of the FWZI (Table 3). The estimated FWZI(H$\beta$) for the 
NLS1 galaxies range from $\sim 6200$ to $\sim 14700$ km\,s$^{-1}$ with
an uncertainty of the order of $\sim 15 - 20$\,\% .

\subsection{Emission-Line Flux Measurements}

Before the measured emission-line fluxes were analyzed, we corrected the 
observed line fluxes for internal reddening. Generally, it is assumed that the 
observed Balmer decrement, i.e. (H$\alpha $/H$\beta) _{obs}$, can be used to 
estimate the reddening, at least for the NLR emission 
(e.g., Davidson \& Netzer 1979; Netzer 1982).
We assumed an intrinsic Balmer decrement of $F(H\alpha)/F(H\beta) _{int} = 2.8$
to estimate $E_{B-V}$ values using the relation

\begin{equation}
{\rm E}_{{\rm B-V}} = {1 \over {{\rm R}(H\beta) - {\rm R}(H\alpha)}} \times
   2.5\,log {F(H\alpha) / F(H\beta) _{obs} \over
             F(H\alpha) / F(H\beta) _{int}}
\end{equation}

\noindent
with R(H$\beta$) and R(H$\alpha$) as corresponding Galactic extinction
coefficients.
Together with a Galactic reddening curve we corrected the observed emission
line ratios relative to F(H$\beta$) applying

\begin{equation}
{F(line) \over F(H\beta)}_{int} = 
  {F(line) \over F(H\beta)}_{obs} \times 
   10^{\,0.4\,\,E_{B-V}\,\,(R_\lambda - 3.68)}
\end{equation}

\noindent
with R$_\lambda$ as the Galactic extinction coefficient at the wavelength of 
the corresponding emission line.
Although there are indications that the intrinsic Balmer decrement is
about $F(H\alpha)/F(H\beta) = 3.1$ under AGN conditions due to collisional
enhancement in the partially ionized zone (e.g., Gaskell \& Ferland 1984; 
Halpern \& Steiner 1983; Netzer 1982) we used the classical value of 2.8
for pure recombination case\,B. 
Assuming case B with $F(H\alpha)/F(H\beta) = 2.8$ appears to be quite
appropriate for the spectra of the studied NLS1s because significant
[\oii ]\,$\lambda 3727$ emission indicates the presence of a large fraction of 
low density gas ($n_e^{crit}(O^+) = 3.1 \times 10^3$ cm$^{-3}$ for $2D_{5/2}$
and $n_e^{crit}(O^+) = 1.6 \times 10^4$ cm$^{-3}$ for $2D_{3/2}$,
Osterbrock 1989). 
Thus, additional collisional enhancement, particularly of H$\alpha$ is not 
expected to be strong.
To estimate the impact of a larger Balmer decrement we also employed 
$F(H\alpha)/F(H\beta) = 3.1$ as intrinsic Balmer decrement. We find that the 
resulting reddening corrected line ratios are less than 10\,\%\ larger for line
ratios shortward of H$\beta$ and less than 10\,\%\ smaller for line ratios 
longward of H$\beta$ using a larger Balmer decrement. Hence, even for emission 
line ratios that involve emission lines like [\oii]\,$\lambda 3727$ and 
[\oiii]\,$\lambda 5007$ the effect is less than 10\,\% . 

As shown in Table 2, we find that the Balmer decrements indicate generally 
larger 
$E_{B-V}$ for the NLR compared to the BLR. However, the $E_{B-V}$ values which
are estimated for the BLR may be overestimated because the intrinsic
H$\alpha$/H$\beta$ ratio may be larger than the value for pure recombination.
Due to collisional enhancement of particularly H$\alpha$ (Gaskell \& Ferland 
1984; Halpern \& Steiner 1983; Netzer 1982) and complex radiative transfer 
effects in the higher density BLR gas, the H$\alpha$/H$\beta$ line ratio might 
be less useful as a reddening indicator (e.g., Davidson \& Netzer 1979).
On average the NLR flux ratios
of $F(H\alpha)$ and $F(H\beta)$ indicate $E_{B-V}(NLR) \simeq 0.^m5 \pm 0.^m1$
compared to $E_{B-V}(BLR) \simeq 0.^m2 \pm 0.^m1$ (Table 2). The average
extinction that we estimate for the NLR is consistent with measurements
given by Wills et al.\,(1993a). 
In Table 4 we provide the observed broad emission-line flux measurements for
the Balmer lines H$\alpha$, H$\beta$, H$\gamma$ and for \hei\,$\lambda 5876$ 
and \heii\,$\lambda 4686$. For the H$\beta $ and \heii\,$\lambda 4686$ emission
lines the observed flux as given by the broad and intermediate broad component
are given separately to illustrate the relative contribution of these 
components to the total broad emission-line flux.
In Table 5, we present the observed and reddening corrected emission-line 
intensities of the NLR emission-line spectrum. The line fluxes 
have been normalized to F(H$\beta $) = 1.0.

\subsection{Diagnostic Emission-Line Ratios}

To specify the nature of the ionizing continuum and to obtain a first estimate 
of the spectral shape we used several emission-line ratios that are suggested 
by Baldwin, Phillips, \& Terlevich (1981) and Veilleux \& Osterbrock (1987). 
In Figure 7 we display the narrow emission-line flux ratios of 
F([\nii]\,$\lambda 6583$)/F(H$\alpha$), 
F([\oi]\,$\lambda 6300$)/F(H$\alpha$), and 
F([\sii]\,$\lambda \lambda 6716,6731$)/F(H$\alpha$) versus 
F([\oiii]\,$\lambda 5007$)/ F(H$\beta$), as well as 
F([\oii]\,$\lambda 3727$)/F([\oiii]\,$\lambda 5007$) versus 
F([\oi]\,$\lambda 6300$)/F(H$\alpha$).
The use of the [\oiii]\,$\lambda 5007$ emission-line profile as a template for
the NLR emission enables a consistent measurement and separation of blended 
narrow emission lines like [\oiii]\,$\lambda 4363$ and the 
[\nii]\,$\lambda 6548,6583$ doublet from H$\gamma$ and H$\alpha$, 
respectively (see Figures 3 and 5).
To illustrate how the NLR emission-line ratios of the NLS1 galaxies relate to 
those of \hii\ regions, starburst galaxies, LINER, and broad-line Seyfert\,1,
we included the results from several other studies in Figure 7.

The location of emission-line ratios that are observed for \hii\ regions
(McCall, Rybski, \& Shields 1985; Veilleux \& Osterbrock 1987; Ho, Filippenko,
\& Sargent 1997) are marked as dots. 
The open triangles are the results for LINERs as given in Ho et al.\,(1997). 
The crosses '$\times$' and '$+$' show 
The location of these line ratios for Seyfert\,1 and Seyfert\,2 galaxies
are shown as ''cross'' and ''plus-sign'' respectively and the are published 
by Koski (1978), Cohen (1983), Veilleux \& Osterbrock (1987), and 
Ho et al.\,(1997). 
As can be seen the line ratios for the NLS1 galaxies (filled diamonds) overlap
well with the region for BLS1 galaxies. The NLS1s are 
clearly separated from the area occupied by \hii\ regions and LINER galaxies.

We find for the NLS1 galaxies under study that the 
F([\oiii]\,$\lambda 5007$)/F(H$\beta _n$) ratio amounts generally to 
$\sim 4$ to $\sim 12$ (Table 5), i.e., a typical range for NLR ratios of 
Seyfert\,1 and Seyfert\,2. It is on average only slightly smaller, which is 
consistent with the result by V\'eron-Cetty et al.\,(2001).
This result also indicates that the narrow H$\beta$ emission-line flux 
cannot be significantly overestimated.
However, two NLS1 galaxies, i.e., Mkn\,291 
and IRAS\,13224-3809, show a low F([\oiii]\,$\lambda 5007$)/F(H$\beta$) ratio 
with $\sim 2.0$ and $\sim 1.1$, respectively, which place them in the range of 
LINER galaxies. 
Together with the low F([\nii]\,$\lambda 6583$)/F(H$\alpha$) ratio these
objects are located in a transition region of Seyfert galaxies and \hii\ 
regions in these diagnostic line ratio diagrams (e.g., Veilleux \& Osterbrock 
1987; Ho et al.\,1997). 

The NLS1 galaxies show similar emission-line ratios like BLS1 galaxies.
It can be seen that low ionization lines like [\oi]\,$\lambda 6300$ in NLS1s
overlap with those in BLS1s, but are slightly weaker on average.
This trend is shown in the F([\oiii]\,$\lambda 5007$)/F(H$\beta$) vs. 
F([\oi]\,$\lambda 6300$)/F(H$\alpha$) diagram.
The F([\oi]\,$\lambda 6300$)/F(H$\alpha$) ratios tend to be shifted to smaller 
ratios than observed for BLS1s.
The tendency of relatively weaker low-ionization lines can be seen even more 
obviously in the F([\oii]\,$\lambda 3727$)/F([\oiii]\,$\lambda 5007$) vs. 
F([\oi]\,$\lambda 6300$)/F(H$\alpha$) diagram (Figure 7). While the
F([\oii]\,$\lambda 3727$)/F([\oiii]\,$\lambda 5007$) ratio covers a similar 
range 
like BLS1 galaxies, i.e., no indications for enhanced [\oii]\,$\lambda 3727$
emission, the F([\oi]\,$\lambda 6300$)/F(H$\alpha$) ratios of the NLS1 galaxies
under study are shifted to lower values. A similar trend can be seen for the 
F([\sii]\,$\lambda 6716,6731$)/F(H$\alpha$) and 
F([\nii]\,$\lambda 6583$)/ F(H$\alpha$) 
ratios. They tend to populate the lower end of the distribution of
these ratios which is shown by BLS1 galaxies (Fig.\,7).
The trend of a little weaker low-ionization line emission on average in these 
NLS1s is consistent with the results of similar studies (e.g.,
Rodriguez-Ardila et al.\,2000).
However, it has also been suggested that NLS1 and BLS1 galaxies show no 
statistically significant difference for the 
F([\oi]\,$\lambda 6300$)/F(H$\alpha$) line ratio (Nagao et al.\,2001).

\subsection{Optical \feii\ Emission --- \feii 4570}

Among the generally observed characteristics of NLS1 galaxies is a strong 
optical \feii\ emission (e.g., Osterbrock \& Pogge 1985; Boller, Brandt, 
\& Fink 1996).
It has been found that the strength of the optical \feii\ emission is 
anti-correlated with the strength of the H$\beta$ emission, the relative
strength of the [\oiii]\,$\lambda 5007$ emission line, as well as with the
profile width of the H$\beta $ emission (e.g., Wills 1982; Gaskell 1985;
Goodrich 1989; Zheng \& O'Brien 1990; Boroson \& Green 1992).

To determine the flux of the optical \feii\ emission we measured the \feii 4570
feature that is dominated by the optical iron multiplets Fe37 and Fe38.
The flux of the \feii 4570 feature is integrated over the range 
$\lambda \lambda 4435 - 4685$\AA\ in the optical \feii\ emission template that 
results from the multi-component fit approach. 
The observed relative \feii 4570 strength, i.e. F(\feii 4570)/F(H$\beta$)
following Gaskell (1985), is shown in Figure 8 as a function of the relative 
[\oiii]\,$\lambda 5007$ emission-line strength relative to H$\beta$. 
On average, the NLS1 galaxies show 
F(\feii 4570)/F(H$\beta$) = $0.76 \pm 0.14$. This result of on average 
relatively larger F(\feii 4570)/F(H$\beta$) ratios is consistent with former 
studies (e.g., Gaskell 1985; Joly 1991 and references therein).
While it is generally assumed that NLS1s exhibit strong \feii\ emission it has 
been suggested that the enhanced F(\feii)/F(H$\beta$) ratios are caused by 
suppressed H$\beta$ emission in higher density gas (Gaskell 1985,2000) in
the case of NLS1 galaxies.

The NLS1 galaxies under study display the well-known trend of a decreasing
\feii 4570 strength for an increasing F([\oiii]\,$\lambda 5007$)/F(H$\beta$)
ratio. However, the trend is dominated by Mkn\,783 and IRAS\,13224-3809
at the low and high end of the F(\feii 4570)/F(H$\beta$) ratios,
respectively (Figure 8).
Although the scatter of the relative \feii 4570 strength is quite 
large we calculated a simple linear fit.
The probability that the correlation occurs by chance amounts to $\sim 15$\%\ 
and hence the trend is only indicated.
At the extreme end of the anti-correlation we find IRAS\,13224-3809 which show 
strong optical \feii\ emission and only relatively weak [\oiii]\,$\lambda 5007$
emission. IRAS\,13224-3809 is a well known extreme strong optical \feii\ 
emitter (e.g., Boller et al.\,1996; V\'eron-Cetty et al.\,2001). The relative 
strength of F(\feii 4570)/F(H$\beta$) = $1.9 \pm 0.3$ is consistent with 
previous results (F(\feii 4570)/F(H$\beta) = 2.4$, Boller et al.\,1996).

\section{Discussion}

\subsection{The BLR Emission-Line Profiles}

Unusually narrow permitted emission lines (${\rm FWHM} \ltsim 2000$ 
km\,s$^{-1}$) are among the characteristic properties that drew attention to 
NLS1 galaxies as an interesting AGN subclass (e.g., Osterbrock \& Pogge 1985;
Goodrich 1989). Later they were found to have distinct X-ray properties, 
particularly rapid variability and a strong soft X-ray excess with a continuum
slope ($F_\nu \propto \nu^{\alpha_x}$) 
$\alpha _x = 2.13 \pm 0.03$ on average compared to 
$\alpha _x \simeq 1.34 \pm 0.03$ for BLS1 galaxies
(e.g., Boller et al.\,1996; Laor et al.\,1997; Grupe et al.\,1998).

As can be seen in Figure 2 the broad permitted H$\beta $ emission-line profile 
show in addition to the characteristic component of intermediate width a strong
broad component as well.
Although the deconvolution into two Gaussian components should be taken
with caution it clearly indicates that high velocity gas is also present in 
NLS1 galaxies.
The width of this broad component of the H$\beta $ profile amounts on average 
to FWHM\,$= 3275 \pm 800$ km\,s$^{-1}$, which is in the typical range of BLS1 
galaxies. For comparison, we measured the profile width of the H$\beta$
emission line for well-known classical BLS1 galaxies, NGC\,3783, NGC\,3516, 
and NGC\,5548. These Seyfert\,1 galaxies show profile widths of 
FWHM(H$\beta$) = 2600, 4000, and 4750 km\,s$^{-1}$, respectively.

The presence of broad profile components in NLS1 galaxies that are typical for 
BLS1 galaxies have already been reported. Rodriguez-Pascual et al.\,(1997) 
detected broad-line emission in several NLS1 galaxies in the ultraviolet for 
Ly$\alpha$, \civ $\lambda 1549$, and \heii\,$\lambda 1640$. But, they did not 
detect broad emission components for the permitted optical lines. They
interpreted this result as an indication that the BLR of NLS1s is lacking 
a significant fraction of partially ionized gas at high velocities and 
hence no broad components for low ionization lines like hydrogen Balmer lines
are detected.
In contrast the presence of a broad component of permitted lines in the 
optical spectra that we detect and that were reported earlier (e.g., 
Grupe et al.\,1999; V\'eron-Cetty et al.\,2001) provides evidence that the 
BLRs of the NLS1s do 
contain a significant fraction of partially ionized gas at high velocities. 
Although the measurement of the FWZI of an emission line depends on the 
assumed continuum level, we estimated the FWZI(H$\beta$) in a range from 
$\sim 6300$ to $\sim 15000$ km\,s$^{-1}$ for the NLS1s of this study. This 
result provides further support for the presence of high velocity low 
ionization gas.
The fraction of the broad emission-line flux that is carried by this broad 
component is on average at least $\sim 60$\%\ of the entire broad emission 
line flux (Table 4).
We conclude that this result indicates that these classical and well-known NLS1
galaxies that we observed do indeed possess a strong broad component that
dominates the broad-line flux. But in contrast to BLS1s these NLS1 
galaxies show a more distinct profile component that is associated with gas of 
intermediate line width, i.e. 
FWHM\,$\simeq 1000 \,{\rm to}\, 2000$ km\,s$^{-1}$.

To compare the relative strength of the broad emission-line flux of NLS1 
galaxies with those of BLS1 galaxies we also measured the equivalent width, 
W$_\lambda$, of the entire H$\beta $ emission-line profile (Table 6).
We find that for the entire broad H$\beta$ profile the equivalent width amounts
on average to W$_\lambda$(H$\beta$) = $72 \pm 29$ \AA\ for the NLS1 galaxies of
our sample. Although this indicates a trend to smaller W$_\lambda$(H$\beta$) 
for NLS1 galaxies, within the uncertainties it is consistent with measurements
of W$_\lambda$(H$\beta$) for BLS1 galaxies of comparable luminosity which is 
quoted to be $\sim 90$ \AA\ (e.g., Osterbrock 1977; Goodrich 1989;
Croom et al.\,2002; Dietrich et al.\,2002).
Although the separation into a broad and an intermediate component 
should be regarded with cautious, we measured W$_\lambda$ for both of these 
components.
We find that the equivalent width of the broad component amounts to 
W$_\lambda$(H$\beta$) = $45 \pm 21$ \AA\ on average. This result suggests that 
the broad component in NLS1 galaxies is about a factor $\sim 2$ weaker than in 
BLS1 galaxies while at the same time the intermediate component is more 
pronounced. 

The composition of the broad emission-line profiles of the studied NLS1s 
is remarkably similar to those found in a small sample of NL\,QSOs. Baldwin 
et al.\,(1988) studied permitted high ionization emission-line profiles, e.g.
\civ $\lambda 1549$. They found that a dominant broad component contains about 
$\sim 80$\,\%\ of the flux. Together with an intermediate component this 
results in narrow composite line profiles. Although Baldwin et al.\,(1988) 
studied high ionization lines, the detection of a similar broad component in 
low ionization lines like H$\beta$ for NLS1 galaxies indicates the presence
of a significant fraction of low ionization gas at high velocity dispersion.
Furthermore, Wills et al.\,(1993b) and Brotherton et al.\,(1994) presented 
models that introduced an intermediate (ILR) and a very broad emission-line 
region (VBLR) to describe the broad emission-line region.
The ILR and VBLR are characterized by profile widths of 
FWHM(ILR)\,$\simeq 2000$ km\,s$^{-1}$ and 
FWHM(VBLR)\,$\simeq 7000$ km\,s$^{-1}$, while the density of the corresponding
regions are $n_H({\rm ILR}) \simeq 10^{10}$ cm$^{-3}$ and 
$n_H({\rm VBLR}) \simeq 10^{12.5}$ cm$^{-3}$. A similar decomposition of the
line profiles has been suggested by Baldwin et al.\,(1996) who studied 
high ionization line profiles of several quasars. They showed that generally 
the broad-line profiles and the observed emission-line ratios can be ascribed 
to three components. They found indications for a narrow component with
FWHM$\simeq 1000$ km\,s$^{-1}$, but in contrast to the ILR this component is 
supposed to have a significantly higher gas density ($n_H \simeq 10^{12.5}$
cm$^{-3}$) than the ILR component.
Very recently, there have been further suggestions that the BLR of active
galactic nuclei can be well described by two distinct emission regions that
can be associated with a broad and intermediate component of the 
observed emission-line profile (e.g., Leighly 2004; Snedden \& Gaskell 2004).

\subsection{The Ionizing EUV -- Soft X-Ray Continuum Shape and Strength}

The diagnostic emission-line ratios that are generally used to distinguish
between different excitation mechanisms, and hence different shapes of the
ionizing continuum, indicate that NLS1 and BLS1 galaxies appear to be very 
similar (Figure 7). To specify the strength of the ionizing continuum and to
explore whether NLS1 galaxies show generally a relatively stronger EUV -- X-ray
continuum, i.e. a more prominent BBB than BLS1 galaxies, the narrow 
emission-line ratio of F(\heii\,$\lambda 4686$)/F(H$\beta$) is a valuable 
tool. This line 
ratio provides information about the relative flux density of ionizing photons 
with energies $\geq 54.4$ eV and $\geq 13.6$ eV, respectively. To trace the EUV
-- soft X-ray continuum to even higher energies we make use of the
F([\nev]\,$\lambda 3426$)/F(H$\beta$) line ratio which yields information at
energies $\geq 97.1$ eV. 
In Figure 9 we show the narrow F(\heii\,$\lambda 4686$)/F(H$\beta$)
versus the F([\nev]\,$\lambda 3426$)/F(H$\beta$) ratio. 
For comparison we included results for BLS1 galaxies taken from the literature
and from former studies which also provided measurements of the 
[\nev]\,$\lambda 3426$ emission line (Koski 1978; Cohen 1983; 
Osterbrock \& Pogge 1985; Kraemer et al.\,1998).
It can be seen that the NLS1 galaxies (filled diamonds) cover a wide range in 
both ratios, the F(\heii\,$\lambda 4686$)/F(H$\beta$) and the 
F([\nev]\,$\lambda 3426$)/F(H$\beta$). 

To estimate a typical ratio for F([\nev]\,$\lambda 3426$) with
F(\heii\,$\lambda 4686$) we computed photoionization models, assuming solar 
abundances (Grevesse \& Anders 1989), for two different
continuum SEDs (Kraemer et al.\,2004). 
The [\nev]\,$\lambda 3426$/\heii\,$\lambda 4686$ will be maximal when the
fraction of Ne$^{+4}$ is in a maximum. For a typical broken power law AGN 
continuum SED, with $F_\nu \propto \nu ^{-\alpha}$, with $\alpha = 1.5$ for 
energies less than 1 keV and $\alpha = 0.8$ for higher energies this occurs
at a ionization parameter U of log\,U = -1.5 and a column density of
$N_H = 10^{21}$ cm$^{-2}$ (matter bounded). The 
[\nev]\,$\lambda 3426$/\heii\,$\lambda 4686$
emission-line ratio that depends also on the gas density can reach values
up to $\sim 5$. The [\nev] emissivity increases as approaching the critical
density for the upper level of the transition ($n_e \simeq 10^7$ cm$^{-3}$;
Osterbrock 1989) while for higher densities collisional de-excitation will
dominate over radiative decay.
If we assume a continuum SED with an additional EUV bump
($\alpha = 0.5$ for energies E\,$<$\,100 eV,
 $\alpha = 2$ for 100\,$<$\,E\,$<$\,1 keV, and 
 $\alpha =0.8$ for E\,$>$\,1 keV) this ratio can reach value up to $\sim 5.5$
Even higher values can be achieved if the continuum is filtered through an
absorber, that is optically thick at the \heii\ Lyman limit.
Most of the F([\nev]\,$\lambda 3426$)/F(H$\beta$) and 
F(\heii\,$\lambda 4686$)/F(H$\beta$) emission-line ratios which be measured 
for the NLS1 galaxies are consistent with the theoretical range that is 
expected for NLR gas photoionized by an AGN continuum as is indicated by the 
dotted lines in Figure 9.
The same trend is
given by the small sample of classical NLS1 galaxies studied by Osterbrock 
\& Pogge (1985). However, 4 NLS1 galaxies of our sample exhibit several
times stronger [\nev]\,$\lambda 3426$ emission relative to H$\beta$ than 
expected
based on the relative strength of the \heii\,$\lambda 4686$ emission line.
To compare their distribution with those of BLS1 galaxies
we include measurements of these line ratios that are observed for BLS1s
(Koski 1978; Cohen 1983; Kraemer et al.\,1999).
In comparison with the NLS1 galaxies the normal BLS1 galaxies show a very
similar distribution of the F(\heii\,$\lambda 4686$)/F(H$\beta$) vs. 
F([\nev]\,$\lambda 3426$)/F(H$\beta$) ratios. While most of the BLS1 galaxies 
also follow the expected relation for these two line ratios like the NLS1s, 
a few 
BLS1 galaxies deviate significantly from this trend. Therefore, NLS1 and BLS1 
galaxies show a very similar distribution of the narrow
F(\heii\,$\lambda 4686$)/F(H$\beta$) and F([\nev]\,$\lambda 3426$)/F(H$\beta$) 
ratios. This indicates that the range of shapes and relative strengths of the 
ionizing EUV continuum in NLS1 and BLS1 galaxies are similar at least up to 
$\sim 100$ eV. 

The similar emission-line ratios that involve low and high ionization lines
and the distribution of these ratios in diagnostic diagrams (Figures 7 and 9)
indicate that the physical conditions of the NLR gas in NLS1 and BLS1
galaxies are comparable.
However, intensive studies on the X-ray continuum shape and strength have
revealed that the soft and hard X-ray spectrum of NLS1 galaxies are generally
steeper compared to those of BLS1s (e.g., Boller et al.\,1996; 
Brandt et al.\,1997; Vaughan et al.\,1999; Leighly 1999a,b). It has also been
found that the parameter $\alpha _{ox}$ that connects the X-ray part of the
spectrum at $\sim 2.5$ keV with the continuum strength at 2500 \AA\ is very
similar for NLS1 and BLS1 galaxies (e.g., Grupe et al.\,1998). This result
is usually interpreted as an indication that NLS1 galaxies show in general 
significant excess emission in the EUV -- soft X-ray range of their continuum. 
However, a recent study by Kraemer et al.\,(2004) detects no strong evidence 
for much stronger soft X-ray -- EUV excess emission in NLS1 galaxies compared 
to those in BLS1 galaxies, investigating the strength of the 2 - 10 keV
X-ray flux with the [\oiii]\,$\lambda 5007$ luminosity for a sample of NLS1
and BLS1 galaxies.

In an upcoming paper we will investigate in more detail the observed emission 
line ratios and the EUV -- soft X-ray excess in the ionizing
continuum. In contrast to modeling the BLR emission spectrum, a detailed study 
of the NLR emission will avoid serious uncertainties of radiative transfer 
aspects that are affecting BLR photoionization models. Alexander et al.\,(1999)
and Kraemer et al.\,(1999,2000) have shown that accurate photoionization 
modeling of the NLR emission provide reliable constraints on the SED of the 
ionizing continuum. 

\section{Conclusion}

We observed a sample of 12 NLS1 galaxies in the optical. To analyze their 
emission-line spectra we
used the [\oiii]\,$\lambda 5007$ emission-line profile as a template to
separate the BLR and NLR flux contributions. NLR emission-line fluxes are
measured for emission lines beginning with [\nev]\,$\lambda 3426$ up to 
[\sii]\,$\lambda 6716,6731$.

We find that the broad emission-line profiles of the permitted lines can be
well represented employing a two Gaussian component fit. The width of the broad
component amounts to ${\rm FWHM} \simeq 3275 \pm 800$ km\,s$^{-1}$ for H$\beta$
which is in the typical range of classical normal BLS1 galaxies. The 
intermediate broad component shows a width of ${\rm FWHM} \simeq 1200 \pm 300$ 
km\,s$^{-1}$. 
Although the approach of using two Gaussian components to describe broad 
emission-line profiles may be oversimplified,
the presence of a broad component with a profile width of several 1000 
km\,s$^{-1}$ FWHM and a FWZI that ranges up to $\sim 15000$ km\,s$^{-1}$
strongly indicates the existence of gas at high velocities in NLS1s.
Furthermore, the emission-line flux that can be associated with the BLR of the 
NLS1 galaxies of our sample amounts to at least 60\%\ of the detected emission 
line flux, i.e. it dominates the total line flux, similar to BLS1 galaxies.

To study the ionizing EUV -- soft X-ray continuum we calculated several narrow
emission-line ratios to trace the shape and strength of the continuum SED.
We find that generally the F([\oiii]\,$\lambda 5007$)/F(H$\beta$) ratio 
varies between $\sim 4$ to 12 for the NLS1 galaxies we studied. Hence, it is 
in the typical range that is observed for BLS1 galaxies with a trend to 
a little smaller values. 

Information about the nature and shape of the ionizing continuum can be 
obtained from diagnostic emission-line ratios. We determined the classical
emission-line flux ratios such as 
F([\nii]\,$\lambda 6583$)/F(H$\alpha$), 
F([\oi]\,$\lambda 6300$)/F(H$\alpha$),\\
F([\sii]\,$\lambda 6716,6731$)/F(H$\alpha$) vs. 
  F([\oiii]\,$\lambda 5007$)/ F(H$\beta$) and
F([\oii]\,$\lambda 3727$)/F([\oiii]\,$\lambda 5007$) vs. 
  F([\oi]\,$\lambda 6300$)/F(H$\alpha$).
The NLS1 galaxies of our sample show a similar distribution of these line
ratios like BLS1 galaxies which reflects that the spectral energy distribution 
of the ionizing EUV -- soft X-ray continuum may not be too different.
Similar to previous studies we find a trend of weaker 
F([\oi]\,$\lambda 6300$)/F(H$\alpha$) emission-line ratios for NLS1s compared 
to BLS1s which can be also seen for the ratios 
F([\nii ]\,$\lambda 6583$)/F(H$\alpha$) and 
F([\sii ]\,$\lambda 6716,6731$)/F(H$\alpha$).
A detected trend of weaker low ionization lines suggests that the NLR gas
of these NLS1 galaxies may be in a state of slightly higher ionization or that
the relative fraction of low ionization gas is reduced.

The relative strength of important forbidden high ionization lines like 
[\nev]\,$\lambda 3426$ compared to narrow \heii\,$\lambda 4686$
provide an additional indication that the ionizing continuum of BLS1 and
NLS1 might be quite similar.
Based on these results we find no strong indications for the presence of a 
strong BBB in the ionizing EUV -- soft X-ray continuum for most of the NLS1 
galaxies of our sample. 
However, the presence of significant EUV -- soft X-ray continuum excess 
emission can be camouflaged by an effective UV absorber which modifies
the ionizing continuum spectrum that excites the NLR gas accordingly.
We will investigate these scenarios using detailed photoionization models 
in more detail in paper II of this study.

\begin{acknowledgements}
      We are grateful to the {\it Small and Moderate Aperture Research 
      Telescope System} (SMARTS) consortium for the observing time at the 
      1.5-m telescope at CTIO to observe this NLS1 galaxy sample. 
\end{acknowledgements}

\clearpage

%
%

\begin{figure*}
\epsscale{2.00}
\plottwo{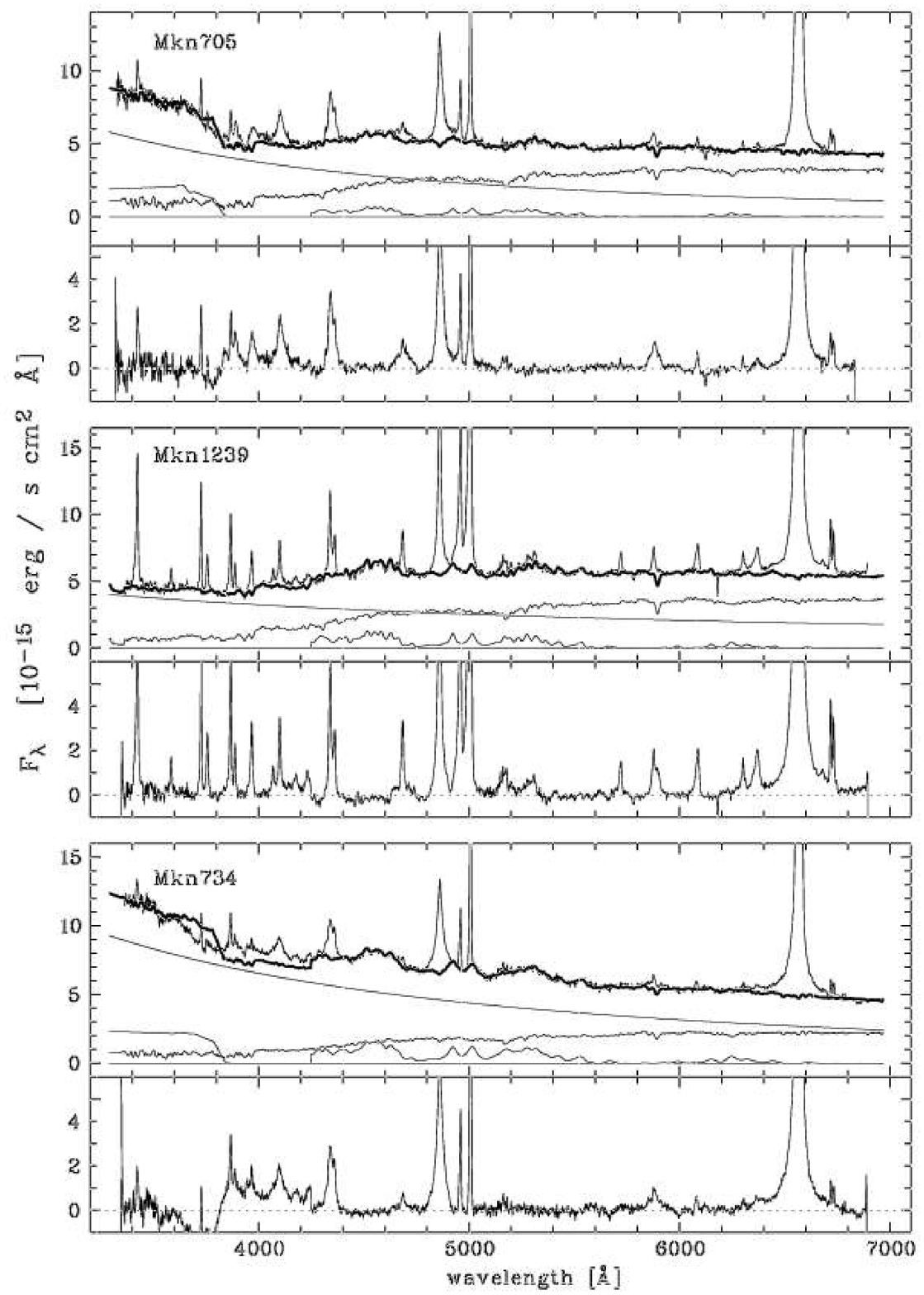}{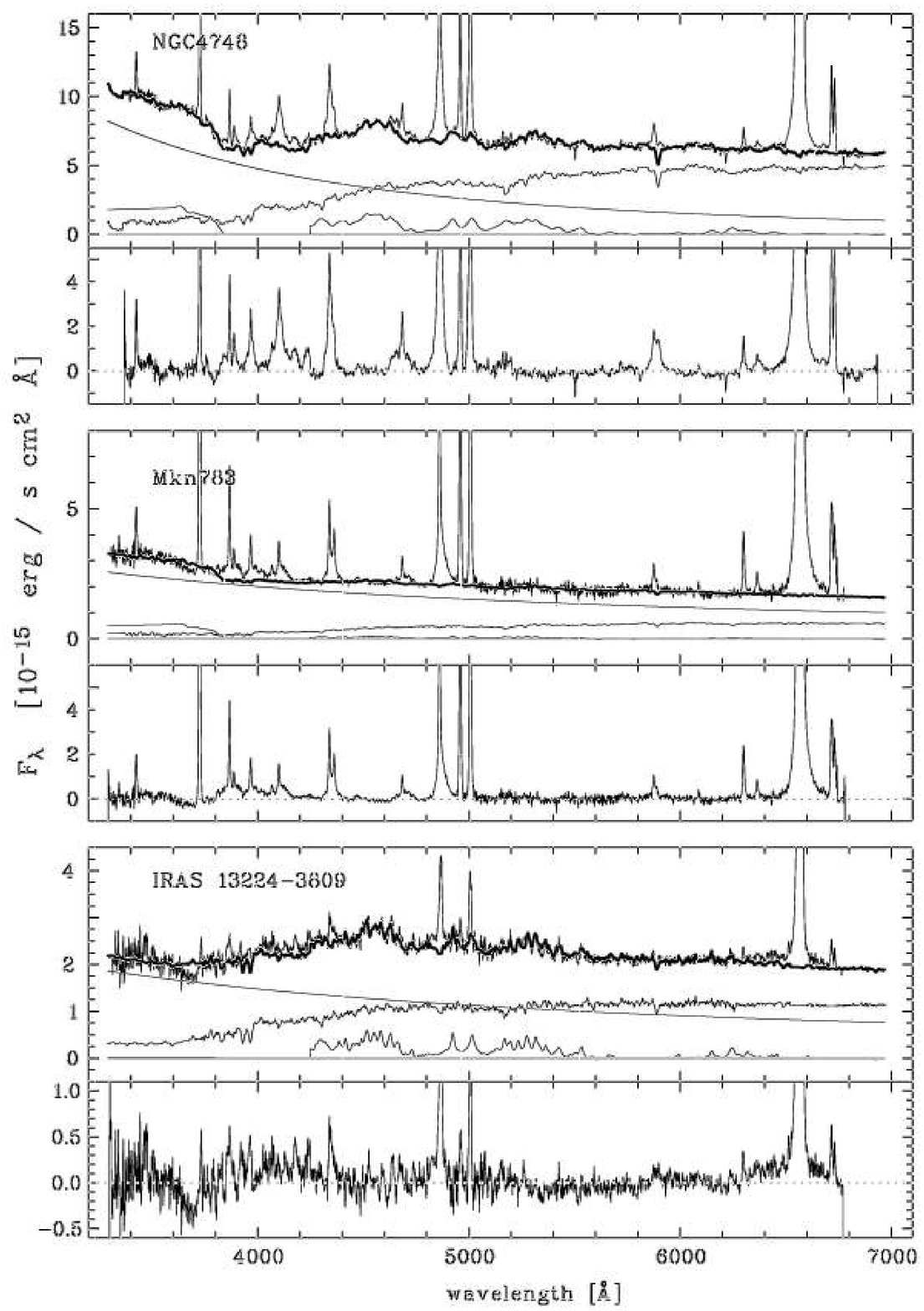}
\label{Figure 1}
\caption{Rest-frame spectra of the NLS1 galaxies -- the flux is 
            given in units of 
            $10^{-15}$ erg\,s$^{-1}$\,cm$^{-2}$\,\AA $^{-1}$. The results of 
            the decomposition of these NLS1 galaxy spectra are displayed.
            The individual components of the multi-component fit approach,
            i.e. a power law continuum, a host galaxy template spectrum, a 
            Balmer continuum spectrum, and an appropriately broadened and
            scaled \feii\ emission spectrum, are shown together with the
            resulting fit (thick line). The residuum spectrum is plotted 
            in the bottom panel of the figures for each galaxy.}
\end{figure*}


\begin{figure*}
\epsscale{2.00}
\plottwo{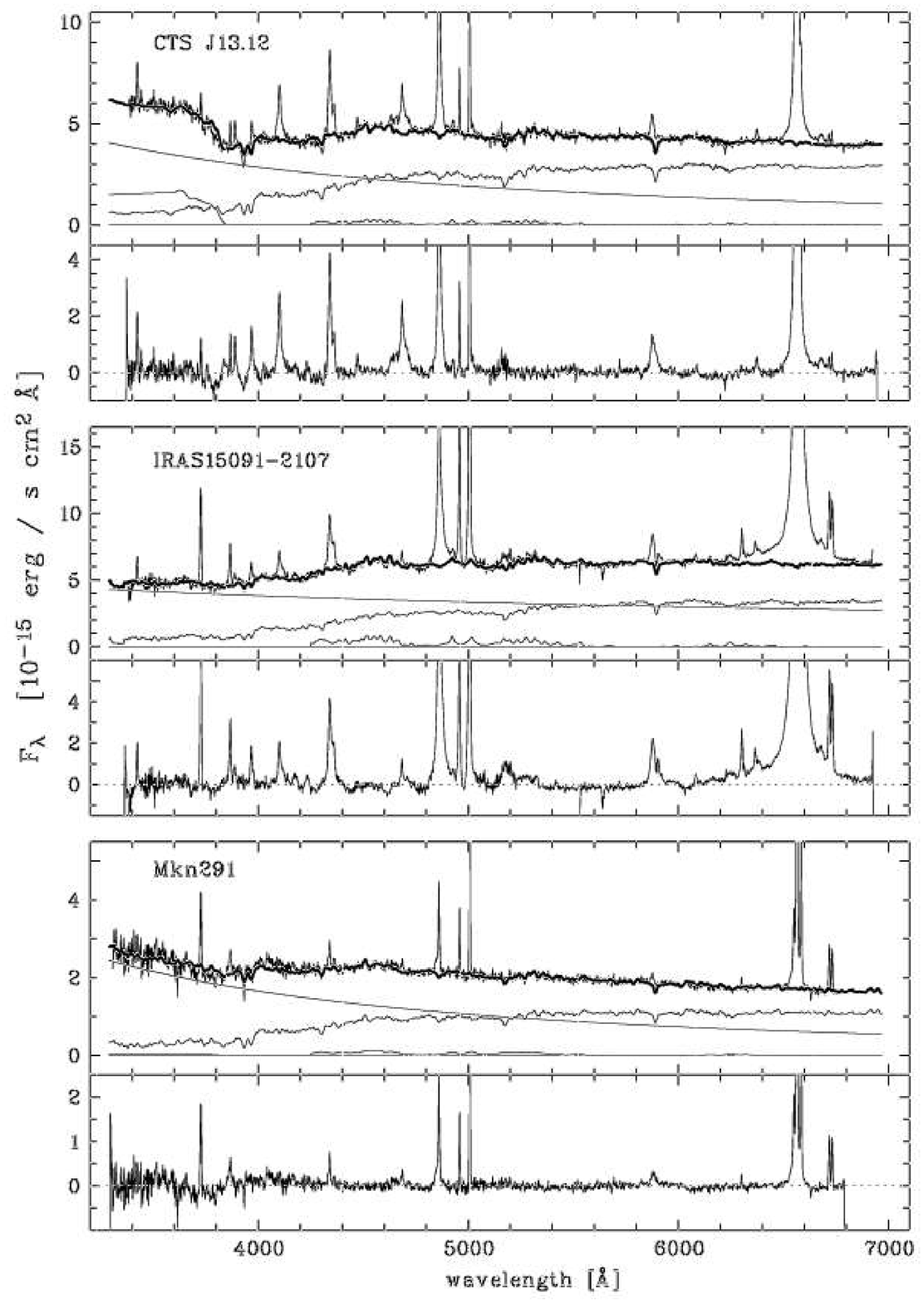}{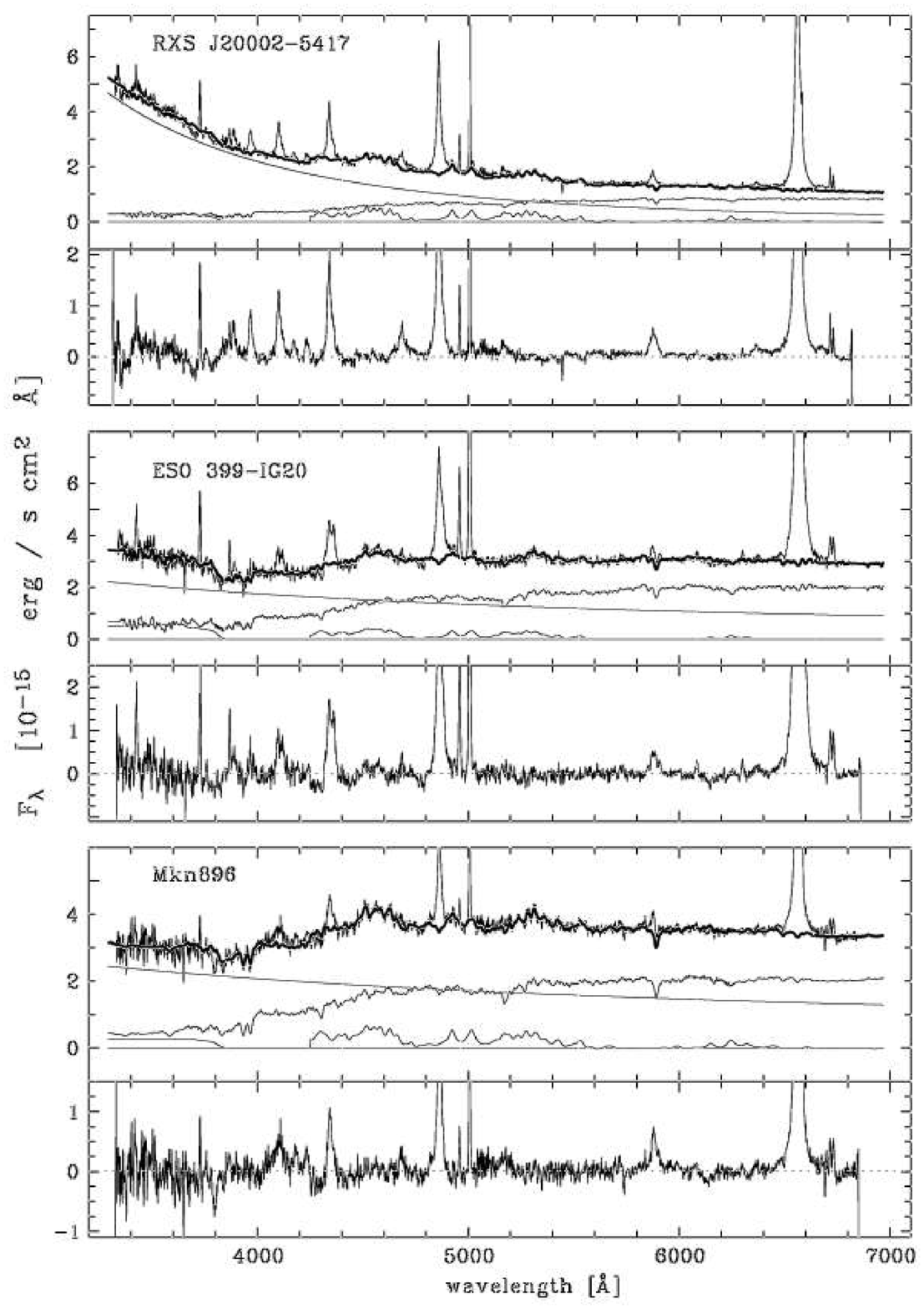}
\label{Figure 1 cont.}

\noindent
Fig.\,1 cont.
\end{figure*}


\begin{figure*}
\epsscale{1.40}
\plotone{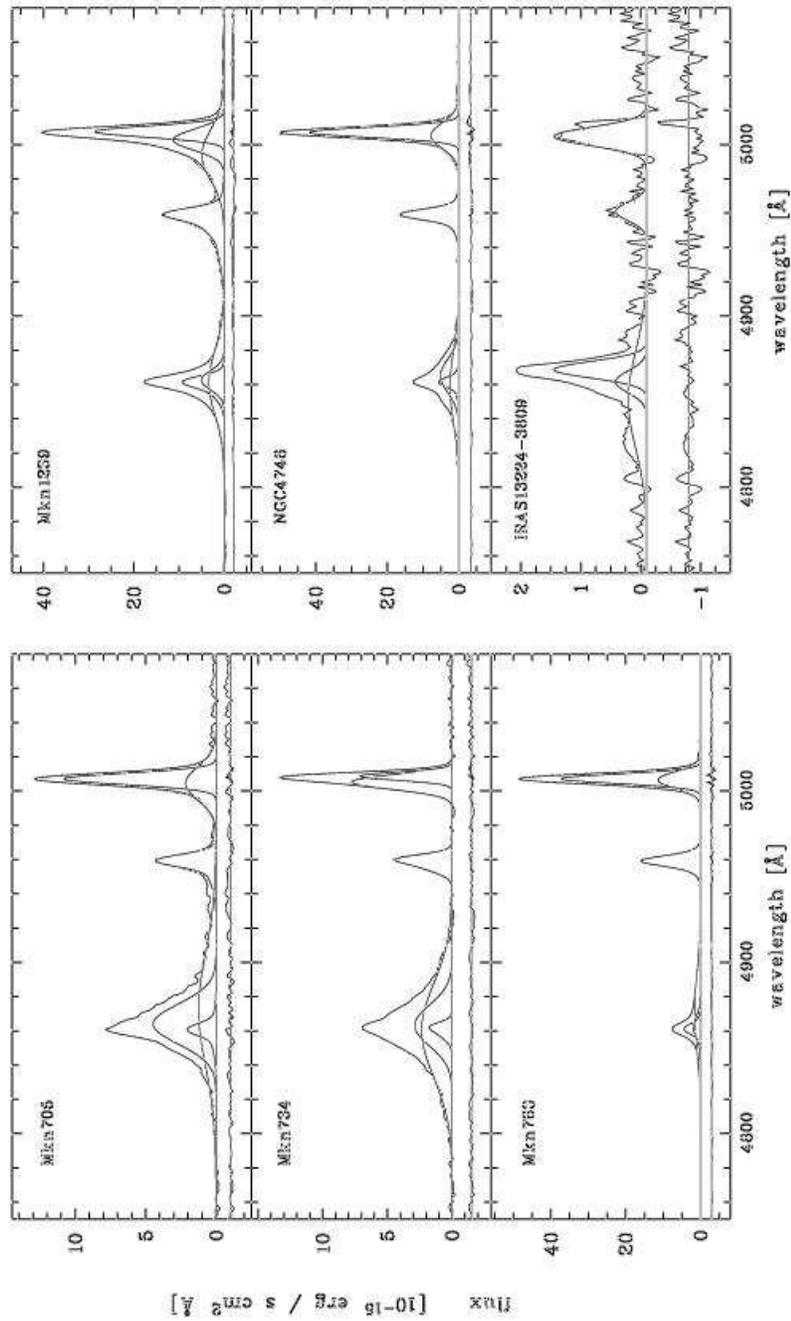}
\label{Figure 2}
\caption{{[}\oiii]\,$\lambda 4959,5007$ and H$\beta$ emission-line 
         profiles.
         For the [\oiii]\,$\lambda 5007$ line profile the individual Gaussian
         components are shown, while for the [\oiii]\,$\lambda 4959$ and
         narrow H$\beta$ line the resulting NLR profile is plotted.
         To illustrate the presence and the relative strength of the broad
         and intermediate broad component both Gaussian profiles are 
         displayed.
         The resulting residuum after subtracting the emission-line profile 
         fits is shown at the bottom of each figure, with a slight offset 
         (indicated by the solid line) to avoid confusion.}
\end{figure*}

\clearpage

\begin{figure*}
\epsscale{1.40}
\plotone{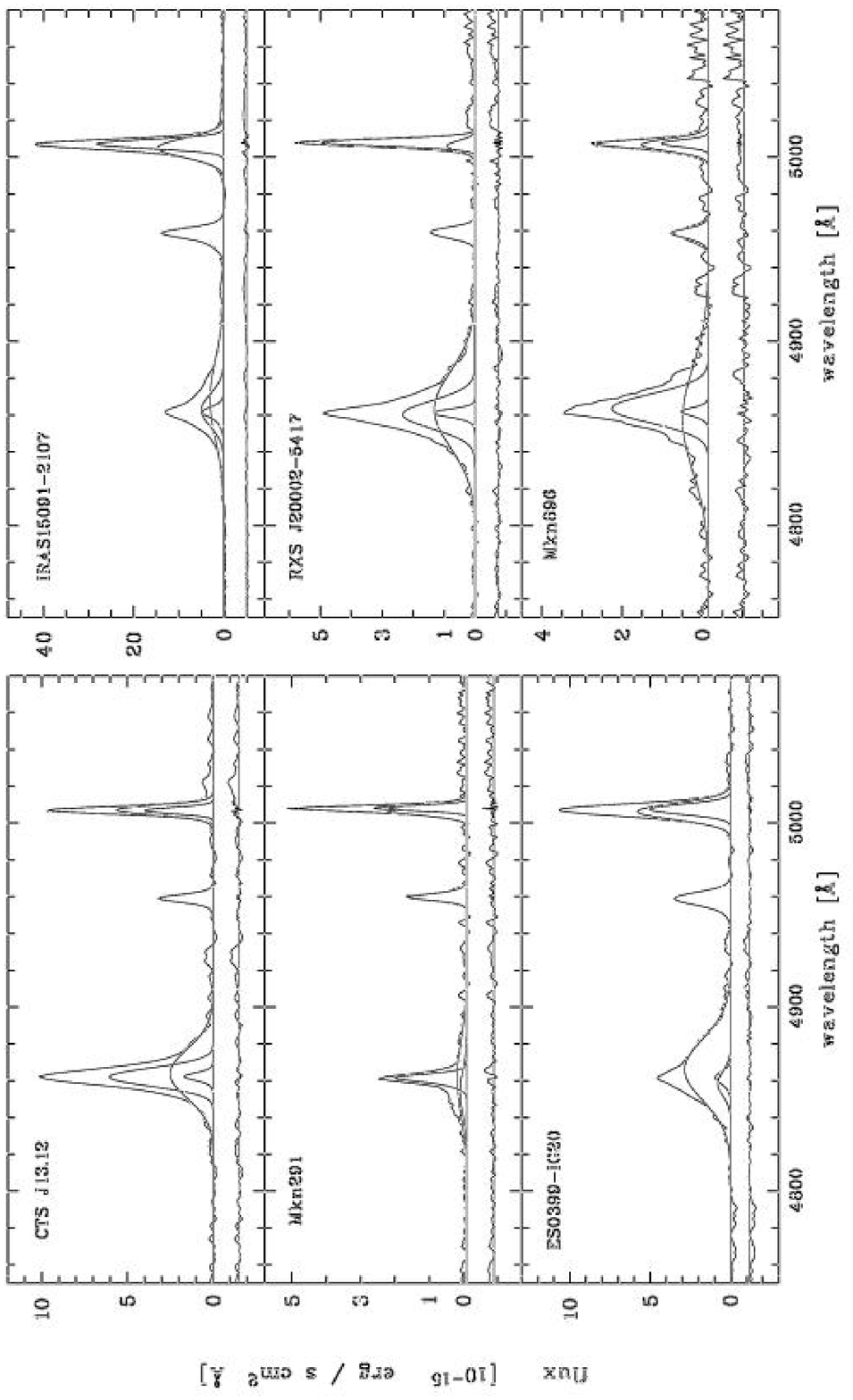}
\label{Figure 2 cont.}

\noindent
Fig.\,2 cont.
\end{figure*}


\begin{figure*}
\epsscale{1.50}
\plotone{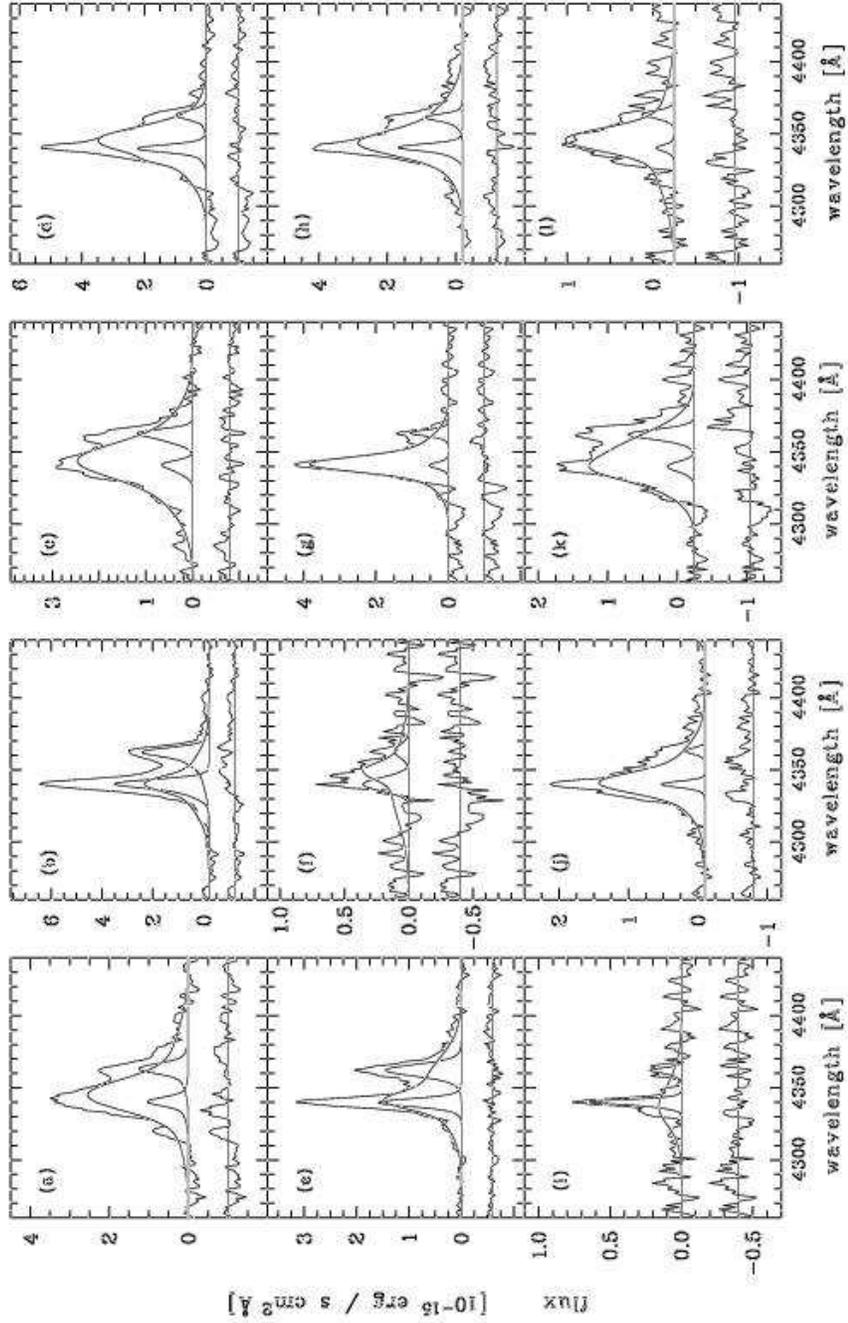}
\label{Figure 3}
\caption{Decomposition of H$\gamma$ -- [\oiii]\,$\lambda 4363$. 
                    The NLS1 galaxies are presented in the following order,
     (a) Mkn\,705, (b) Mkn\,1239, (c) Mkn\,734. (d) NGC\,4748,
     (e) Mkn\,783, (f) IRAS\,13224-3809, (g) CTS\,J13.12, (h) IRAS\,15091-2107,
     (i) Mkn\,291, (j) RSX\,J20002-5417, (k) ESO\,399IG20, (l) Mkn\,896.}
\end{figure*}


\begin{figure*}
\epsscale{1.50}
\plotone{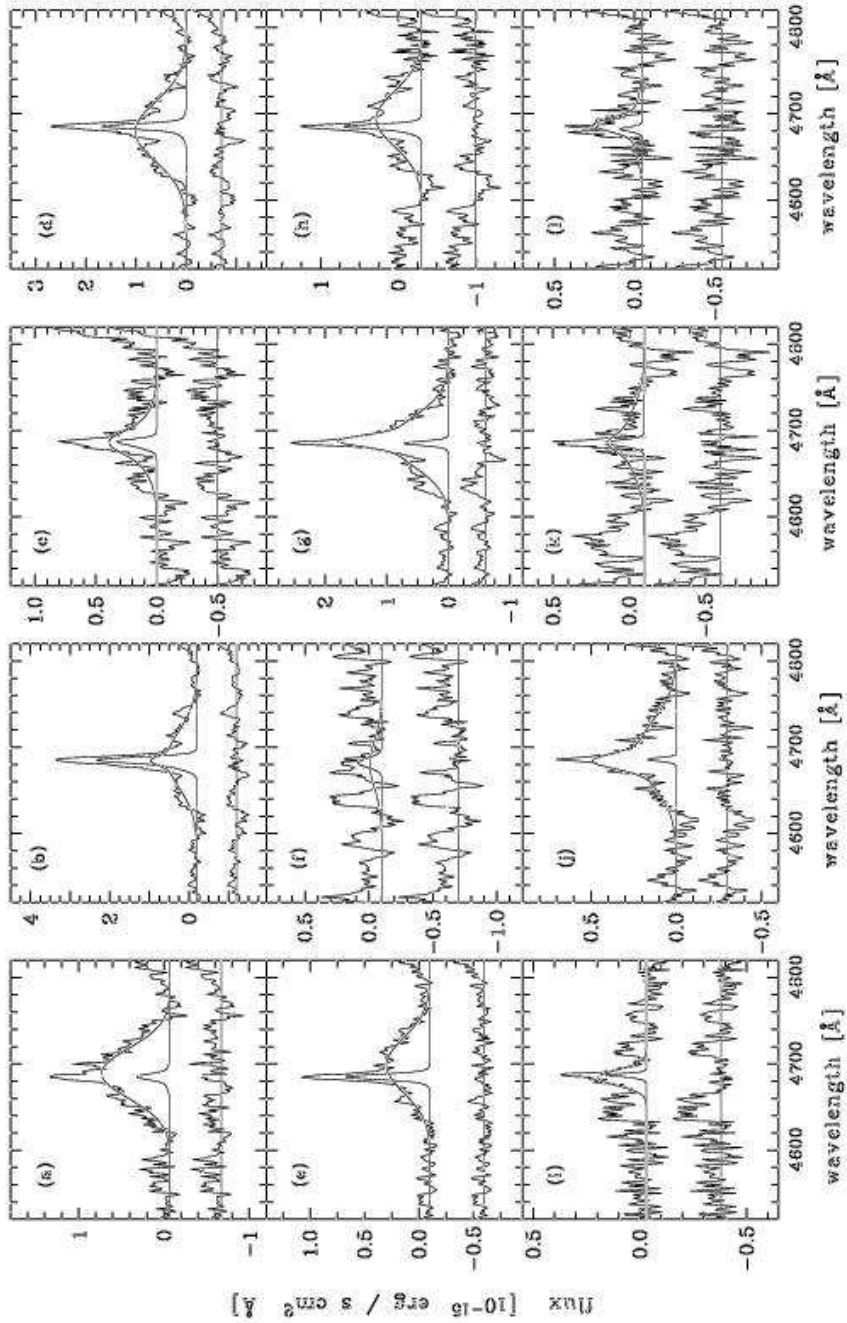}
\label{Figure 4}
\caption{Same as Figure 3; decomposition of the 
                    \heii\,$\lambda 4686$ emission-line profile. The NLS1 
                    galaxies are in the same order as in Figure 3.}
\end{figure*}


\begin{figure*}
\epsscale{1.50}
\plotone{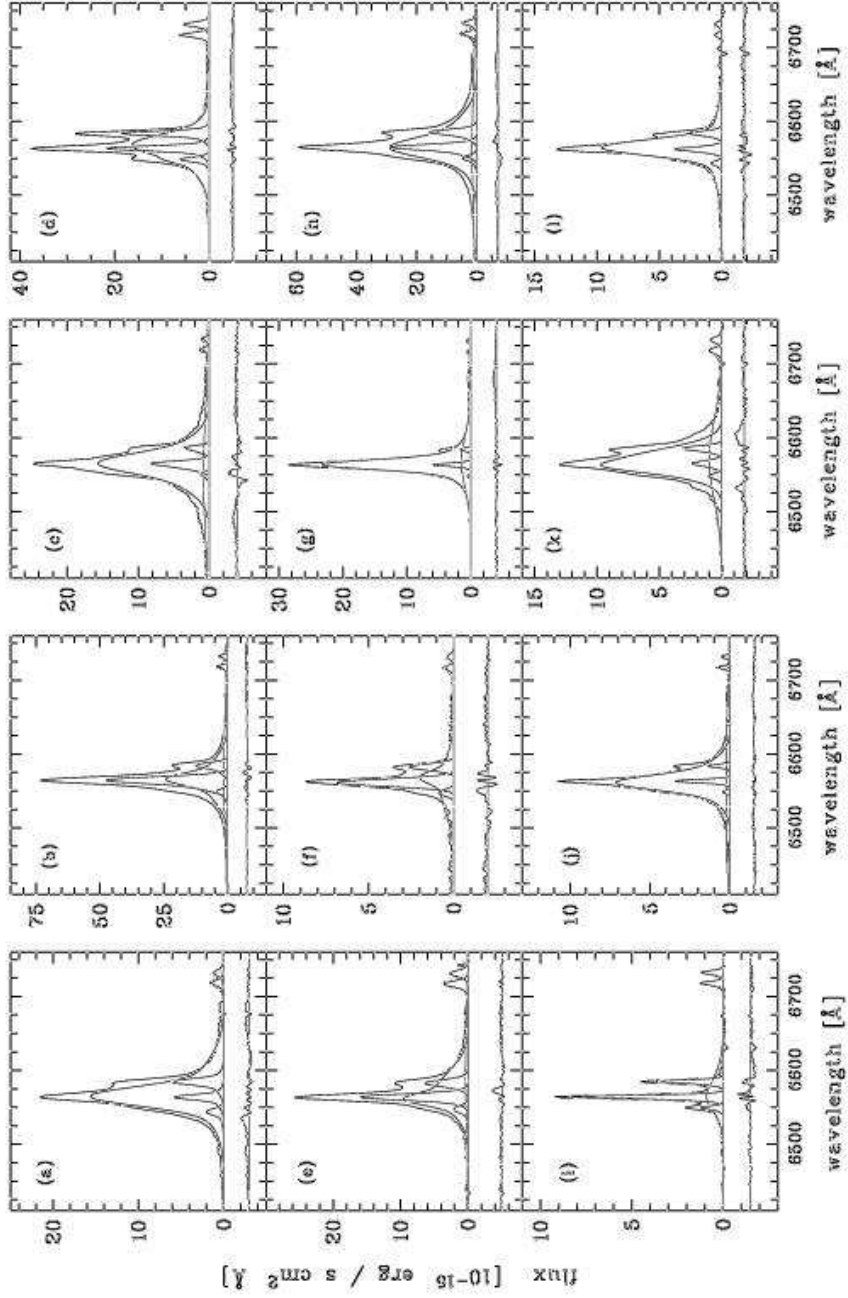}
\label{Figure 5}
\caption{Same as Figure 3; decomposition of the 
           H$\alpha$ and [\nii]\,$\lambda 6548,6583$ emission-line profiles is
           displayed. The NLS1 galaxies are in the same order as in Figure 3.}
\end{figure*}


\begin{figure*}
\epsscale{1.00}
\plotone{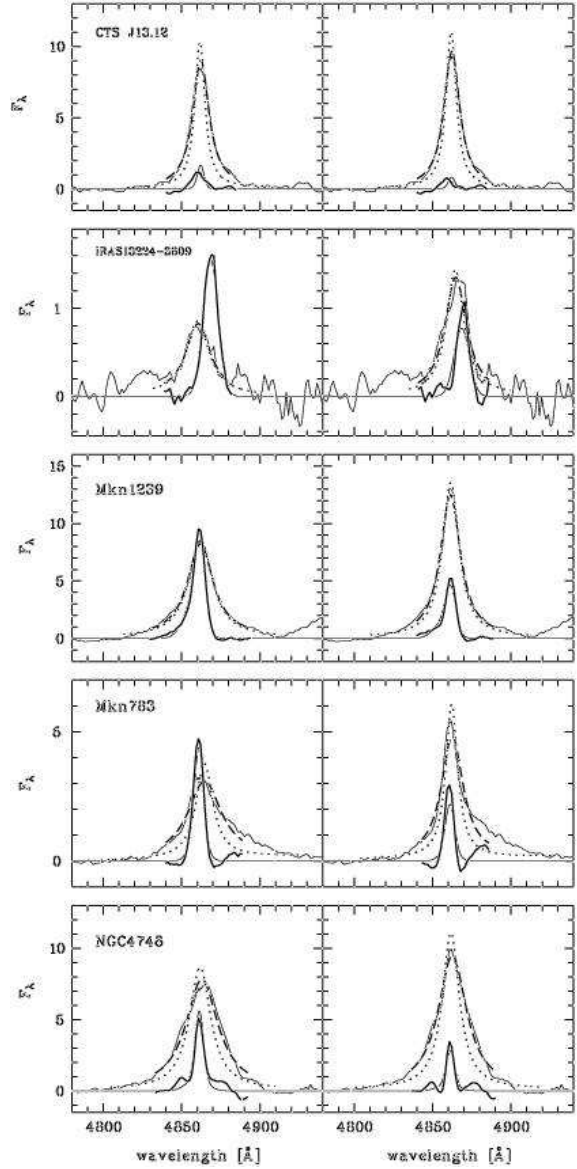}
\label{Figure 6}
\caption{Typical examples for the Lorentzian profile fits we
           obtained for the broad H$\beta$ emission-line profile.
           In the left panel the residuum is shown after subtracting a narrow 
           H$\beta$ component that is as strong as it has been determined.
           In the right panel the corresponding residuum is displayed assuming
           that the narrow H$\beta$ component is 50\,\%\ less intense. The
           Lorentzian profile fits that are limited to the inner part of the 
           residuum are plotted as short dashed line, while the fit for the 
           entire residuum, i.e. including the profile wings is shown as
           dotted line. The resulting narrow residuum (thick line) which could
           be associated with the narrow H$\beta$ line flux is presented for
           comparison with a properly scaled narrow emission profile template.}
\end{figure*}


\begin{figure*}
\epsscale{2.10}
\plotone{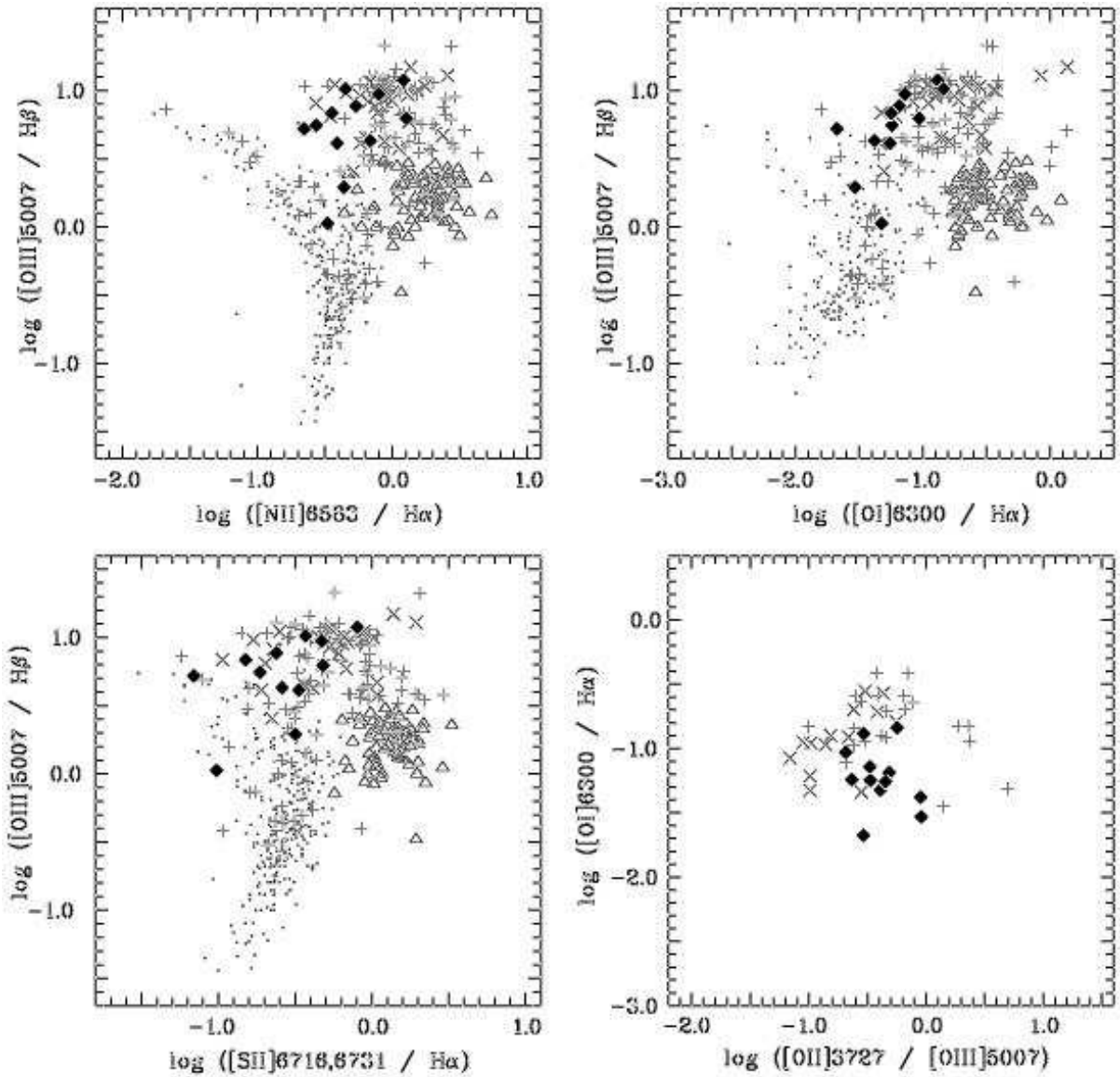}
\label{Figure 7}
\caption{Diagnostic emission-line ratios to compare the
            NLS1 galaxies under study (filled diamonds) with the typical ratios
            that are measured for the NLR emission of normal Seyfert\,1 
            (crosses) and 2 galaxies ($+$), LINER (open triangles), and 
            \hii\ regions (dots) as available in the literature
            (Koski 1978; Cohen 1983; McCall, Rybski, \& Shields 1985; 
            Veilleux \& Osterbrock 1987; Ho, Filippenko, \& Sargent 1997).
            Clockwise beginning with the upper left panel 
            (i) log F([\nii]\,$\lambda 6583$)/F(H$\alpha$) vs.
                log F([\oiii]\,$\lambda 5007$)/F(H$\beta$),
           (ii) log F([\oi]\,$\lambda 6300$)/F(H$\alpha$) vs.
                log F([\oiii]\,$\lambda 5007$)/F(H$\beta$),   
          (iii) log F([\sii]\,$\lambda 6716,6731$)/F(H$\alpha$) vs.
                log F([\oiii]\,$\lambda 5007$)/F(H$\beta$),
           (iv) log F([\oii]\,$\lambda 3727$)/F([\oiii]\,$\lambda 5007$) vs.
                log F([\oi]\,$\lambda 6300$)/F(H$\alpha$).}
\end{figure*}


\begin{figure*}
\plotone{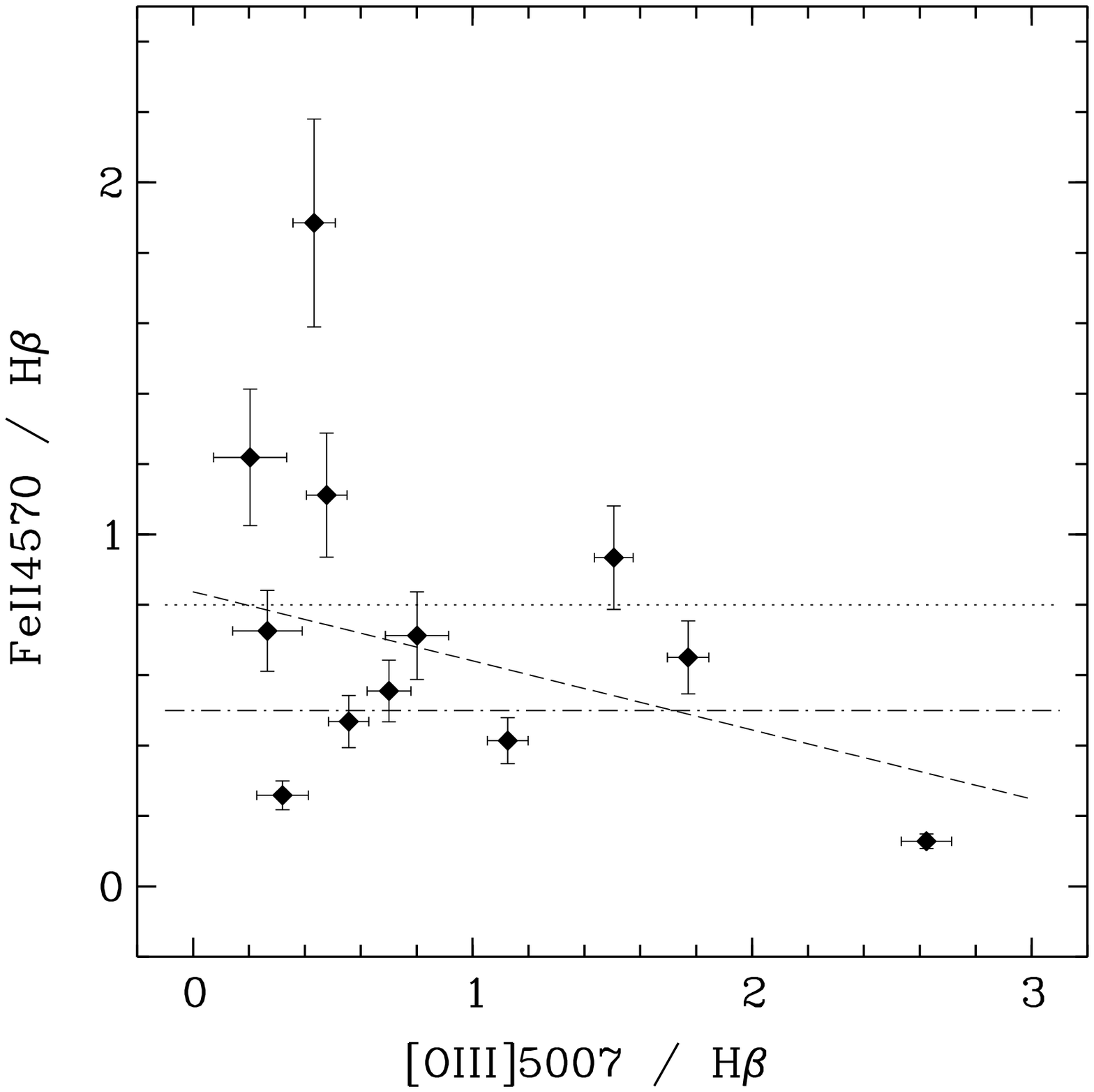}
\label{Figure 8}
\caption{The relative optical \feii\ emission-line strength as 
            represented by the F(\feii 4570)/F(H$\beta$) emission-line ratio as
            a function of F([\oiii]\,$\lambda 5007$)/F(H$\beta$).
            An anti-correlation of increasing an 
            F(\feii 4570)/F(H$\beta$) ratio can be seen for decreasing 
            F([\oiii]\,$\lambda 5007$)/F(H$\beta$).
            The dashed - dotted line indicates the mean 
            F(\feii 4570)/F(H$\beta$) ratio for BLS1 galaxies while the dotted 
            line marks a typical value for the F(\feii 4570)/F(H$\beta$) that 
            is measured for NLS1 galaxies (e.g., Joly 1991 and references 
            therein).}
\end{figure*}


\begin{figure*}
\plotone{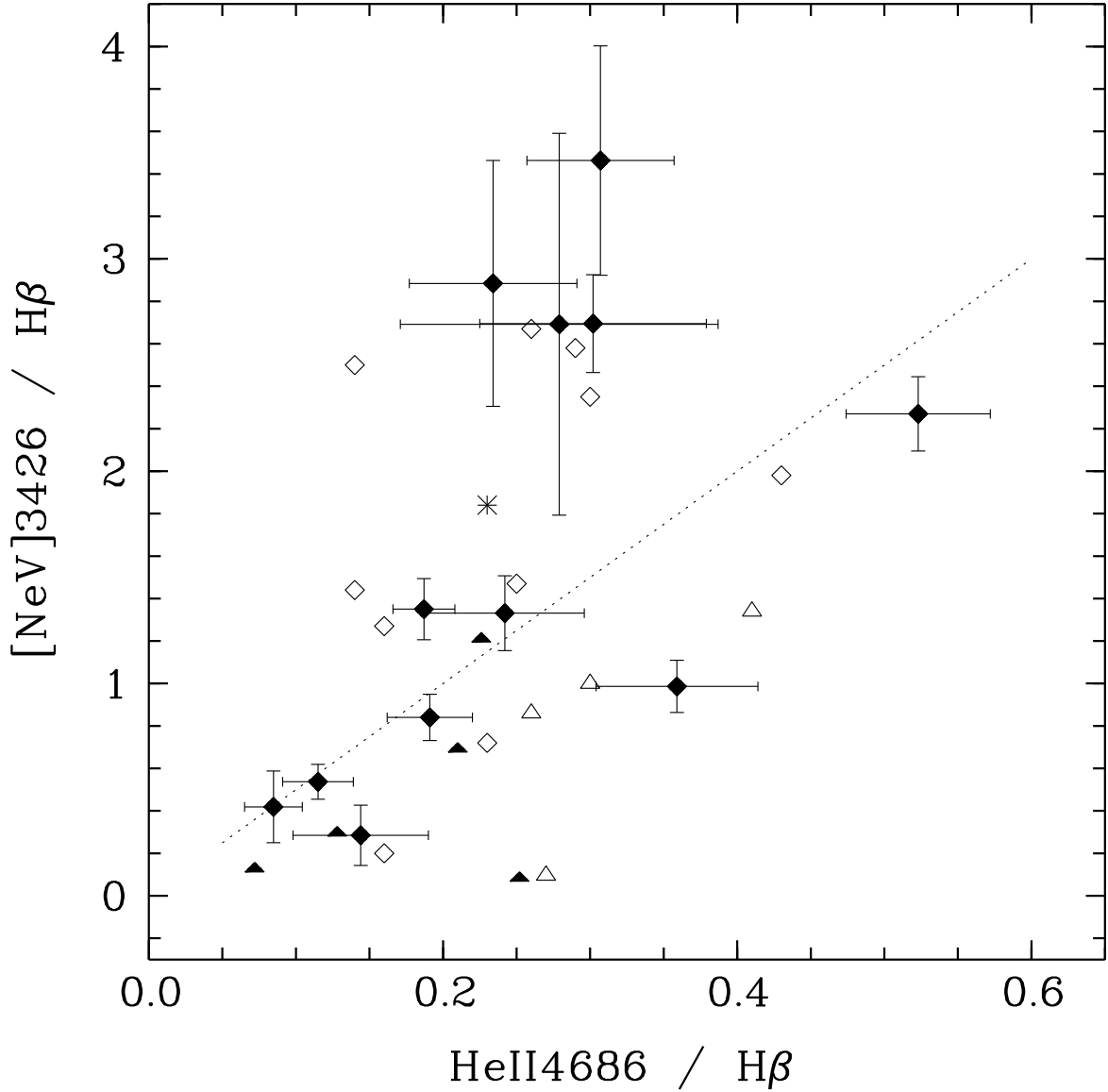}
\label{Figure 9}
\caption{Comparison of the narrow emission-line ratios 
            F(\heii\,$\lambda 4686$)/F(H$\beta$) vs. 
            F([\nev]\,$\lambda3426$)/F(H$\beta$)
            which we measure for the NLS1 galaxies of our sample (filled 
            diamonds) with measurements provided by Osterbrock \& Pogge (1985)
            (filled triangles) and for normal Seyfert\,1 Galaxies (Koski 1978)
            (open diamonds); Cohen (1983) (open triangles); Kraemer et 
            al.\,(1998) (asterix). 
            The dotted line indicate the expected value of 
            $5 \times$ F(\heii\,$\lambda 4686$)/F(H$\beta$) $\simeq$ 
            F([\nev]\,$\lambda3426$)/F(H$\beta$)
            (based on photoionization models, paper II).}

\end{figure*}


\hspace*{-15mm}
\begin{deluxetable}{lcccccc}
\tablewidth{0pt}
\tabletypesize{\scriptsize}
\tablecaption{The Narrow-Line Seyfert 1 Galaxy sample.}
\tablehead{
\colhead{object} &
\colhead{RA\,$^a$}&
\colhead{DEC\,$^a$}&
\colhead{m$_v$}&
\colhead{z\,$^b$}&
\colhead{$E_{B-V}$\,$^c$}&
\colhead{$\Gamma _X$\,$^d$}\\
\colhead{(1)} &
\colhead{(2)} &
\colhead{(3)} &
\colhead{(4)} &
\colhead{(5)} &
\colhead{(6)} &
\colhead{(7)}
}
\startdata
Mkn\,705        &09 26 03.3&$+$12 44 03&14.60&0.028&0.058&$-2.53\pm0.09 ^2$\\
Mkn\,1239       &09 52 19.1&$-$01 36 44&14.49&0.019&0.067&$-4.06\pm0.30 ^2$\\
Mkn\,734        &11 21 47.1&$+$11 44 19&14.93&0.049&0.048&$-3.63\pm0.19 ^2$\\
NGC\,4748       &12 52 12.4&$-$13 24 53&14.03&0.014&0.061& \nodata \\
Mkn\,783        &13 02 58.9&$+$16 24 28&15.55&0.067&0.033&$-1.30\pm0.72 ^2$\\
IRAS\,13224-3809&13 25 19.2&$-$38 24 54&13.80&0.065&0.080&$-4.4 \pm0.2  ^1$\\
CTS J13.12      &13 51 29.4&$-$18 13 47&15.50&0.012&0.129& \nodata \\
IRAS\,15091-2107&15 11 59.8&$-$21 19 02&14.83&0.044&0.142&$-2.61\pm1.04 ^2$\\
Mkn\,291        &15 55 08.0&$+$19 11 33&15.51&0.035&0.058&$-3.41\pm0.79 ^2$\\
RXS\,J20002-5417&20 00 15.4&$-$54 17 12&16.50&0.061&0.084& \nodata \\
ESO\,399-IG20   &20 06 58.1&$-$34 32 55&14.51&0.025&0.118& \nodata \\
Mkn\,896        &20 46 20.8&$-$02 48 45&14.61&0.027&0.083&$-3.38\pm0.05 ^2$\\
\enddata
\tablenotetext{a}{right ascension and declination in J2000.0}
\tablenotetext{b}{based on the [\oiii ]\,$\lambda 4959,5007$ emission-line 
                  profiles}
\tablenotetext{c}{Galactic extinction, taken from Dickey \& Lockman (1990)}
\tablenotetext{d}{ROSAT power-law fit photon index $\Gamma _X$ of the 
                  soft X-ray spectrum\\ 
          \hspace*{5mm}(1: Boller et al.\,1996; 2: Pfefferkorn et al.\,2001).}
\end{deluxetable}

\hspace*{-15mm}
\begin{deluxetable}{lccccc}
\tablewidth{0pt}
\tabletypesize{\scriptsize}
\tablecaption{Results of the multi-component spectral fit.}
\tablehead{
\colhead{object}     &
\colhead{$\alpha$\,$^a$}   &
\colhead{Hubble-type\,$^b$}&
\colhead{rel.F$_{gal}$\,$^c$} &
\colhead{E$_{B-V}$\,$^d$}  &
\colhead{E$_{B-V}$\,$^d$}  \\
\colhead{   } &
\colhead{   } &
\colhead{   } &
\colhead{   } &
\colhead{NLR} &
\colhead{BLR} \\
\colhead{(1)} &
\colhead{(2)} &
\colhead{(3)} &
\colhead{(4)} &
\colhead{(5)} &
\colhead{(6)}
}
\startdata
Mkn\,705        &+0.25&S0& 51 \% & 0.01& 0.05\\
Mkn\,1239       &-0.90&Sb& 48 \% & 0.84& 0.32\\
Mkn\,734        &-0.20&S0& 27 \% & 0.66& 0.07\\
NGC\,4748       &+0.80&Sb& 56 \% & 0.48& 0.06\\
Mkn\,783        &-0.75&S0& 23 \% & 0.54& 0.22\\
IRAS\,13224-3809&-0.80&Sb& 45 \% & 0.46& 0.11\\
CTS J13.12      &-0.20&E & 56 \% & 0.25& 0.23\\
IRAS\,15091-2107&-1.40&Sb& 42 \% & 1.01& 0.54\\
Mkn\,291        & 0.00&Sa& 47 \% & 0.40& 0.00\\
RXS\,J20002-5417&+1.90&S0& 38 \% &-0.06\,$^e$&-0.26\,$^e$\\
ESO\,399-IG20   &-0.80&S0& 51 \% & 0.05& 0.00\\
Mkn\,896        &-1.15&E & 46 \% & 0.99& 0.20\\
\enddata
\tablenotetext{a}{Spectral slope $\alpha$, $F_\nu \propto \nu^\alpha$}
\tablenotetext{b}{Hubble type as indicated by the multi-component fit}
\tablenotetext{c}{relative flux of the host galaxy fit to the total flux 
                  observed at $\lambda = 5100$ \AA .}
\tablenotetext{d}{estimated reddening E$_{B-V}$ assuming
                  F(H$\alpha$)/F(H$\beta)$ = 2.8}
\tablenotetext{e}{no reddening correction was applied if $E_{B-V}$ is less
                  than zero}
\end{deluxetable}

\hspace*{-15mm}
\begin{deluxetable}{lccccccc}
\tablewidth{0pt}
\tabletypesize{\scriptsize}
\tablecaption{Decomposition of the broad H$\beta$ and \heii\,$\lambda 4686$ 
              emission-line profiles.}
\tablehead{
\colhead{object} &
\colhead{ } &
\colhead{FWHM(H$\beta$)\,$^a$} &
\colhead{ } & 
\colhead{FWZI(H$\beta$)\,$^a$} &
\colhead{ } &
\colhead{FWHM(\heii )\,$^a$} &
\colhead{ } \\
\colhead{ } &
\colhead{broad}  &
\colhead{interm.}&
\colhead{narrow} &
\colhead{ }      &
\colhead{broad}  &
\colhead{interm.}&
\colhead{narrow}\\
\colhead{(1)} &
\colhead{(2)} &
\colhead{(3)} &
\colhead{(4)} & 
\colhead{(5)} &
\colhead{(6)} &
\colhead{(7)} &
\colhead{(8)} 
}
\startdata
Mkn\,705        &$4780 \pm 240$&$1687 \pm 84$&$543 \pm 19$&$14750\pm 1850$&$4758 \pm 2.62$& \nodata     &$564 \pm 28$\\
Mkn\,1239       &$2733 \pm 136$&$ 808 \pm 48$&$564 \pm 31$&$ 7530\pm  960$&$4982 \pm 2.39$&$ 791 \pm 29$&$592 \pm 32$\\
Mkn\,734        &$3941 \pm 200$&$1732 \pm 87$&$544 \pm 27$&$12740\pm 2760$&$4089 \pm 3.32$&$1797 \pm 85$&$646 \pm 32$\\
NGC\,4748       &$3050 \pm 152$&$1403 \pm 70$&$421 \pm 19$&$ 9880\pm 1440$&$5153 \pm 2.67$& \nodata     &$491 \pm 25$\\
Mkn\,783        &$3205 \pm 160$&$ 900 \pm 45$&$454 \pm 25$&$10060\pm 1610$&$4764 \pm 3.90$& \nodata     &$471 \pm 24$\\
IRAS\,13224-3809&$4044 \pm 206$&$ 913 \pm 46$&$700 \pm 31$&$11035\pm 2300$&$4196 \pm 2.20$& \nodata     &$816 \pm 39$\\
CTS J13.12      &$1930 \pm  95$&$ 722 \pm 36$&$311 \pm 16$&$ 6295\pm  710$&$4365 \pm 2.35$&$ 927 \pm 46$&$379 \pm 19$\\
IRAS\,15091-2107&$3158 \pm 158$&$1278 \pm 64$&$441 \pm 25$&$10430\pm 1510$&$4728 \pm 2.36$& \nodata     &$457 \pm 23$\\
Mkn\,291        &$2944 \pm 147$&$1646 \pm 95$&$322 \pm 12$&$ 8020\pm 1590$&  \nodata      &$1381 \pm 69$&$292 \pm 19$\\
RXS\,J20002-5417&$2992 \pm 150$&$1093 \pm 55$&$326 \pm 25$&$ 9315\pm 1390$&$5575 \pm 3.13$&$1268 \pm 63$&$305 \pm 21$\\
ESO\,399-IG20   &$2425 \pm 121$&$ 944 \pm 47$&$479 \pm 24$&$ 7960\pm 1170$&$5053 \pm 2.79$&$1370 \pm 69$&$535 \pm 26$\\
Mkn\,896        &$4098 \pm 205$&$1125 \pm 56$&$325 \pm 18$&$12035\pm 2090$&$4251 \pm 2.97$&$1168 \pm 58$&$396 \pm 21$\\
\enddata
\tablenotetext{a}{Full width at half maximum (FWHM) and full width at zero
                  intensity (FWZI) are given in units of km\,s$^{-1}$.}
\end{deluxetable}

\hspace*{-15mm}
\begin{deluxetable}{lccccccc}
\tablewidth{0pt}
\tabletypesize{\scriptsize}
\tablecaption{Emission-line flux measurements of the observed broad and 
              intermediate H$\beta$ and \heii\,$\lambda 4686$ emission-line
              profiles, as well as the total broad-line flux of H$\alpha$, 
              H$\gamma$, and \hei $\lambda 5876$ (in units of 
              $10^{-15}$ erg\,s$^{-1}$\,cm$^{-2}$).}
\tablehead{
\colhead{object}  &
\colhead{H$\beta$}&
\colhead{H$\beta$}& 
\colhead{\heii   }&
\colhead{\heii   }&
\colhead{H$\alpha$}& 
\colhead{H$\gamma$}& 
\colhead{\hei }\\
\colhead{ }       &
\colhead{broad}   &
\colhead{interm.} &
\colhead{broad}   &
\colhead{interm.} &
\colhead{total}       &
\colhead{total}       &
\colhead{total}       \\
\colhead{(1)} &
\colhead{(2)} &
\colhead{(3)} &
\colhead{(4)} & 
\colhead{(5)} & 
\colhead{(6)} &
\colhead{(7)} &
\colhead{(8)} 
}
\startdata
Mkn\,705        &$102.4 \pm 4.3$&$130.1 \pm 1.7$&$ 58.2 \pm 2.8$& \nodata      &$684.8\pm 69.0$&$93.4\pm 11.2$&$48.1\pm 0.7$\\
Mkn\,1239       &$159.8 \pm 2.1$&$ 70.2 \pm 0.8$&$ 61.8 \pm 1.7$&$ 5.6 \pm 0.6$&$895.8\pm 48.0$&$61.6\pm  2.0$&$39.5\pm 4.0$\\
Mkn\,734        &$159.9 \pm 4.5$&$ 86.3 \pm 2.3$&$ 12.0 \pm 0.7$&$ 6.5 \pm 0.4$&$738.6\pm 40.0$&$103.9\pm 7.9$&$44.8\pm 7.0$\\
NGC\,4748       &$127.4 \pm 4.6$&$123.8 \pm 2.6$&$ 87.1 \pm 1.4$& \nodata      &$760.8\pm 52.5$&$107.5\pm 11.5$&$59.1\pm 6.5$\\
Mkn\,783        &$ 87.3 \pm 1.1$&$ 28.9 \pm 0.5$&$ 31.7 \pm 0.5$& \nodata      &$409.1\pm 32.0$&$46.8\pm  3.6$&$18.4\pm 0.4$\\
IRAS\,13224-3809&$ 21.5 \pm 1.4$&$  8.2 \pm 0.3$&$  6.5 \pm 2.0$& \nodata      &$ 93.7\pm 10.0$&$11.7\pm  2.6$&$ 7.9\pm 2.3$\\
CTS J13.12      &$ 83.5 \pm 2.1$&$ 75.4 \pm 0.9$&$ 67.3 \pm 2.4$&$13.4 \pm 0.7$&$567.3\pm 35.0$&$71.5\pm  1.1$&$37.4\pm 2.6$\\
IRAS\,15091-2107&$167.6 \pm 5.7$&$109.8 \pm 2.8$&$ 46.3 \pm 1.1$& \nodata      &$1367\pm 140  $&$95.4\pm  2.0$&$76.3\pm 1.2$\\
Mkn\,291        &$ 13.9 \pm 0.5$&$  5.5 \pm 0.5$&  \nodata      &$ 5.2 \pm 0.2$&$ 54.5\pm  8.0$&$ 6.1\pm  1.1$&$ 7.9\pm 0.2$\\
RXS\,J20002-5417&$ 68.1 \pm 1.7$&$ 44.0 \pm 1.1$&$ 21.1 \pm 1.0$&$ 5.7 \pm 0.3$&$240.5\pm 25.0$&$42.0\pm  3.6$&$19.1\pm 0.2$\\
ESO\,399-IG20   &$119.5 \pm 1.4$&$ 13.0 \pm 0.7$&$ 11.4 \pm 1.0$&$ 2.4 \pm 0.4$&$371.0\pm 48.1$&$58.3\pm  5.4$&$20.1\pm 1.0$\\
Mkn\,896        &$ 45.4 \pm 1.9$&$ 46.8 \pm 0.7$&$  4.5 \pm 0.4$&$ 4.7 \pm 0.2$&$319.3\pm 33.5$&$38.7\pm  3.9$&$23.7\pm 2.3$\\
\enddata
\end{deluxetable}


\begin{table*}
{\footnotesize
\caption{NLR emission-line ratios relative to F(H$\beta$).\label{tbl-5}}
\begin{tabular}{lcccccc}
\tableline\tableline
line  &\multicolumn{2}{c}{Mkn705} & \multicolumn{2}{c}{Mkn1239}& \multicolumn{2}{c}{Mkn734}\\
      &obs.&corr.&obs.&corr.&obs.&corr.\\
(1)   & (2) & (3) & (4) & (5) & (6) & (7) \\
\tableline
{[}\nev ]3346  &  \nodata       &  \nodata       &  \nodata       &  \nodata       &  \nodata       &  \nodata       \\
{[}\nev ]3426  &$1.34\pm 0.14$  &$1.35\pm 0.14$  &$1.32\pm 0.20$  &$3.46\pm 0.54$  &$1.35\pm 0.27$  &$2.88\pm 0.58$  \\
{[}\fevii ]3588&  \nodata       &  \nodata       &$0.23\pm 0.06$  &$0.53\pm 0.13$  &  \nodata       &  \nodata       \\
{[}\oii ]3727  &$1.28\pm 0.13$  &$1.29\pm 0.13$  &$0.72\pm 0.09$  &$1.53\pm 0.19$  &$1.27\pm 0.19$  &$2.29\pm 0.34$  \\
{[}\fevii ]3760&$0.62\pm 0.07$  &$0.63\pm 0.07$  &$0.38\pm 0.06$  &$0.78\pm 0.12$  &  \nodata       &  \nodata       \\
{[}\neiii ]3869&$0.79\pm 0.18$  &$0.80\pm 0.18$  &$0.67\pm 0.11$  &$1.29\pm 0.21$  &$1.26\pm 0.16$  &$2.10\pm 0.27$  \\
HeI3889        &$0.17\pm 0.03$  &$0.17\pm 0.03$  &$0.12\pm 0.04$  &$0.23\pm 0.07$  &$0.084\pm 0.025$&$0.14\pm 0.04$  \\
H$\zeta$3888   &$0.088\pm 0.017$&$0.088\pm 0.017$&$0.033\pm 0.010$&$0.063\pm 0.018$&$0.096\pm 0.031$&$0.16\pm 0.05$  \\
{[}\neiii ]3968&$0.20\pm 0.03$  &$0.20\pm 0.03$  &$0.22\pm 0.04$  &$0.40\pm 0.07$  &$0.33\pm 0.07$  &$0.53\pm 0.11$  \\
H$\epsilon$3970&$0.12\pm 0.02$  &$0.12\pm 0.02$  &$0.065\pm 0.014$&$0.12\pm 0.02$  &$0.059\pm 0.012$&$0.093\pm 0.019$\\
{[}\sii ]4072  &  \nodata       &  \nodata       &$0.14\pm 0.02$  &$0.23\pm 0.04$  &$0.20\pm 0.04$  &$0.30\pm 0.07$  \\
H$\delta$ 4101 &$0.23\pm 0.04$  &$0.23\pm 0.04$  &$0.13\pm 0.02$  &$0.22\pm 0.03$  &$0.13\pm 0.02$  &$0.19\pm 0.03$  \\
H$\gamma$ 4340 &$0.44\pm 0.06$  &$0.44\pm 0.06$  &$0.35\pm 0.05$  &$0.50\pm 0.06$  &$0.37\pm 0.06$  &$0.49\pm 0.08$  \\
{[}\oiii ]4363 &$0.61\pm 0.08$  &$0.61\pm 0.08$  &$0.33\pm 0.05$  &$0.46\pm 0.07$  &$0.67\pm 0.11$  &$0.88\pm 0.14$  \\
HeI4471        &  \nodata       &  \nodata       &$0.024\pm 0.011$&$0.032\pm 0.015$&  \nodata       &  \nodata       \\
HeII4686       &$0.19\pm 0.02$  &$0.19\pm 0.02$  &$0.27\pm 0.04$  &$0.31\pm 0.05$  &$0.21\pm 0.05$  &$0.23\pm 0.06$  \\
H$\beta$ 4861  &$1.00\pm 0.07$  &$1.00\pm 0.07$  &$1.00\pm 0.10$  &$1.00\pm 0.10$  &$1.00\pm 0.10$  &$1.00\pm 0.10$  \\
{[}\oiii ]4959 &$2.08\pm 0.15$  &$2.08\pm 0.15$  &$1.93\pm 0.20$  &$1.81\pm 0.19$  &$2.47\pm 0.28$  &$2.34\pm 0.26$  \\
{[}\oiii ]5007 &$6.25\pm 0.45$  &$6.25\pm 0.45$  &$5.80\pm 0.58$  &$5.24\pm 0.52$  &$7.41\pm 0.77$  &$6.84\pm 0.71$  \\
{[}\fevii ]5159&$0.24\pm 0.04$  &$0.24\pm 0.04$  &$0.15\pm 0.02$  &$0.12\pm 0.02$  &$0.46\pm 0.09$  &$0.39\pm 0.07$  \\
{[}\fevi ]5176 &$0.26\pm 0.04$  &$0.26\pm 0.04$  &$0.17\pm 0.03$  &$0.14\pm 0.03$  &  \nodata       &  \nodata       \\
{[}\Ni ]5199   &$0.088\pm 0.019$&$0.088\pm 0.019$&$0.072\pm 0.013$&$0.057\pm 0.011$&  \nodata       &   \nodata      \\
{[}\fevii ]5278&  \nodata       &  \nodata       &$0.10\pm 0.02$  &$0.076\pm 0.014$&  \nodata       &  \nodata       \\
{[}\fevii ]5283&  \nodata       &  \nodata       &  \nodata       &  \nodata       &  \nodata       &  \nodata       \\
{[}\fexiv ]5307&  \nodata       &  \nodata       &$0.096\pm 0.018$&$0.070\pm 0.013$&  \nodata       &  \nodata       \\
{[}\cav ]5309  &$0.094\pm 0.023$&$0.094\pm 0.023$&$0.076\pm 0.014$&$0.055\pm 0.010$&  \nodata       &  \nodata       \\
{[}\fevii ]5721&$0.19\pm 0.04$  &$0.19\pm 0.04$  &$0.27\pm 0.04$  &$0.15\pm 0.02$  &$0.18\pm 0.03$  &$0.12\pm 0.02$  \\
{[}\nii ]5755  &  \nodata       &  \nodata       &  \nodata       &  \nodata       &$0.11\pm 0.03$  &$0.07\pm 0.02$  \\
HeI5876        &$0.18\pm 0.04$  &$0.18\pm 0.04$  &$0.17\pm 0.03$  &$0.092\pm 0.014$&$0.17\pm 0.05$  &$0.11\pm 0.03$  \\
{[}\fevii ]6087&$0.34\pm 0.05$  &$0.34\pm 0.05$  &$0.42\pm 0.06$  &$0.21\pm 0.03$  &$0.51\pm 0.09$  &$0.30\pm 0.05$  \\
{[}\oi ]6300   &$0.26\pm 0.03$  &$0.26\pm 0.03$  &$0.13\pm 0.02$  &$0.059\pm 0.008$&$0.29\pm 0.07$  &$0.16\pm 0.04$  \\
{[}\siii ]6312 &$0.058\pm 0.010$&$0.057\pm 0.010$&$0.028\pm 0.006$&$0.013\pm 0.003$&$0.11\pm 0.03$  &$0.06\pm 0.02$  \\
{[}\oi ]6364   &$0.088\pm 0.017$&$0.087\pm 0.017$&$0.043\pm 0.007$&$0.019\pm 0.003$&$0.097\pm 0.026$&$0.052\pm 0.014$\\
{[}\fex ]6374  &$0.43\pm 0.06$  &$0.43\pm 0.06$  &$0.16\pm 0.03$  &$0.072\pm 0.013$&  \nodata       &  \nodata       \\
{[}\nii ]6548  &$1.18\pm 0.22$  &$1.18\pm 0.22$  &$0.51\pm 0.09$  &$0.21\pm 0.04$  &$0.67\pm 0.17$  &$0.34\pm 0.09$  \\
H$\alpha$ 6563 &$2.81\pm 0.40$  &$2.80\pm 0.40$  &$6.74\pm 0.80$  &$2.79\pm 0.33$  &$5.58\pm 0.93$  &$2.79\pm 0.47$  \\
{[}\nii ]6583  &$3.57\pm 0.63$  &$3.55\pm 0.63$  &$1.50\pm 0.22$  &$0.62\pm 0.09$  &$2.00\pm 0.30$  &$1.00\pm 0.15$  \\
HeI6678        &  \nodata       &  \nodata       &$0.053\pm 0.008$&$0.021\pm 0.003$&  \nodata       &  \nodata       \\
{[}\sii ]6716  &$0.73\pm 0.07$  &$0.72\pm 0.07$  &$0.27\pm 0.03$  &$0.11\pm 0.01$  &$0.47\pm 0.09$  &$0.23\pm 0.04$  \\
{[}\sii ]6731  &$0.62\pm 0.07$  &$0.62\pm 0.07$  &$0.23\pm 0.03$  &$0.087\pm 0.011$&$0.41\pm 0.07$  &$0.19\pm 0.04$  \\
\tableline
\end{tabular}
}
\end{table*}

\begin{table*}
\setcounter{table}{4}
{\footnotesize
\caption{cont.\label{tbl-5}}
\begin{tabular}{lcccccc}
\tableline\tableline
line  &\multicolumn{2}{c}{NGC4748} & \multicolumn{2}{c}{Mkn783}
& \multicolumn{2}{c}{IRAS13224-3809}\\
      &obs.&corr.&obs.&corr.&obs.&corr.\\
(1)   & (2) & (3) & (4) & (5) & (6) & (7) \\
\tableline
{[}\nev ]3346  &  \nodata       &   \nodata      &$0.14\pm 0.02$  &$0.29\pm 0.05$  &  \nodata       &  \nodata       \\
{[}\nev ]3426  &$0.57\pm 0.07$  &$0.99\pm 0.12$  &$0.45\pm 0.06$  &$0.84\pm 0.11$  &$0.17\pm 0.08$  &$0.29\pm 0.14$  \\
{[}\fevii ]3588&  \nodata       &   \nodata      &  \nodata       &  \nodata       &  \nodata       &  \nodata       \\
{[}\oii ]3727  &$2.02\pm 0.21$  &$3.11\pm 0.33$  &$3.59\pm 0.40$  &$5.81\pm 0.65$  &$0.29\pm 0.03$  &$0.43\pm 0.05$  \\
{[}\fevii ]3760&$0.18\pm 0.03$  &$0.28\pm 0.04$  &$0.055\pm 0.022$&$0.088\pm 0.036$&  \nodata       &  \nodata       \\
{[}\neiii ]3869&$0.73\pm 0.10$  &$1.06\pm 0.14$  &$0.89\pm 0.11$  &$1.35\pm 0.17$  &$0.26\pm 0.05$  &$0.38\pm 0.09$  \\
HeI3889        &$0.10\pm 0.03$  &$0.14\pm 0.04$  &$0.039\pm 0.014$&$0.06\pm 0.02$  &  \nodata       &  \nodata       \\
H$\zeta$3888   &$0.07\pm 0.02$  &$0.10\pm 0.03$  &$0.052\pm 0.020$&$0.08\pm 0.03$  &  \nodata       &  \nodata       \\
{[}\neiii ]3968&$0.22\pm 0.04$  &$0.31\pm 0.06$  &$0.20\pm 0.03$  &$0.29\pm 0.05$  &  \nodata       &  \nodata       \\
H$\epsilon$3970&$0.093\pm 0.015$&$0.13\pm 0.02$  &$0.078\pm 0.022$&$0.11\pm 0.03$  &  \nodata       &  \nodata       \\
{[}\sii ]4072  &$0.06\pm 0.01$  &$0.084\pm0.014$ &$0.096\pm 0.019$&$0.13\pm 0.03$  &  \nodata       &  \nodata       \\
H$\delta$ 4101 &$0.15\pm 0.03$  &$0.21\pm 0.03$  &$0.14\pm 0.03$  &$0.20\pm 0.03$  &  \nodata       &  \nodata       \\
H$\gamma$ 4340 &$0.37\pm 0.05$  &$0.45\pm 0.06$  &$0.32\pm 0.05$  &$0.41\pm 0.07$  &$0.22\pm 0.06$  &$0.27\pm 0.07$  \\
{[}\oiii ]4363 &$0.20\pm 0.02$  &$0.24\pm 0.02$  &$0.44\pm 0.05$  &$0.55\pm 0.07$  &$0.067\pm 0.016$&$0.08\pm 0.02$  \\
HeI4471        &$0.036\pm 0.013$&$0.042\pm 0.013$&$0.008\pm 0.003$&$0.010\pm 0.003$&  \nodata       &  \nodata       \\
HeII4686       &$0.33\pm 0.05$  &$0.36\pm 0.06$  &$0.18\pm 0.03$  &$0.19\pm 0.03$  &$0.13\pm 0.04$  &$0.14\pm 0.05$  \\
H$\beta$ 4861  &$1.00\pm 0.08$  &$1.00\pm 0.08$  &$1.00\pm 0.08$  &$1.00\pm 0.08$  &$1.00\pm 0.04$  &$1.00\pm 0.04$  \\
{[}\oiii ]4959 &$3.28\pm 0.28$  &$3.15\pm 0.27$  &$3.68\pm 0.50$  &$3.52\pm 0.48$  &$0.37\pm 0.02$  &$0.36\pm 0.02$  \\
{[}\oiii ]5007 &$9.97\pm 0.79$  &$9.41\pm 0.75$  &$10.98\pm 1.03$ &$10.28\pm 0.96$ &$1.12\pm 0.06$  &$1.06\pm 0.06$  \\
{[}\fevii ]5159&$0.18\pm 0.03$  &$0.16\pm 0.02$  &$0.055\pm 0.025$&$0.05\pm 0.02$  &$0.14\pm 0.04$  &$0.13\pm 0.03$  \\
{[}\fevi ]5176 &$0.16\pm 0.02$  &$0.14\pm 0.02$  &  \nodata       & \nodata        &  \nodata       &  \nodata       \\
{[}\Ni ]5199   &$0.12\pm 0.02$  &$0.11\pm 0.02$  &  \nodata       & \nodata        &  \nodata       &  \nodata       \\
{[}\fevii ]5278&  \nodata       &  \nodata       &  \nodata       & \nodata        &  \nodata       &  \nodata       \\
{[}\fevii ]5283&  \nodata       &  \nodata       &  \nodata       & \nodata        &  \nodata       &  \nodata       \\
{[}\fexiv ]5307&$0.074\pm 0.015$&$0.062\pm 0.013$&  \nodata       & \nodata        &  \nodata       &  \nodata       \\
{[}\cav ]5309  &  \nodata       &  \nodata       &  \nodata       & \nodata        &  \nodata       &  \nodata       \\
{[}\fevii ]5721&$0.094\pm 0.015$&$0.07\pm 0.01$  &$0.033\pm 0.017$&$0.023\pm 0.012$&  \nodata       &  \nodata       \\
{[}\nii ]5755  &  \nodata       &  \nodata       &  \nodata       &  \nodata       &  \nodata       &  \nodata       \\
HeI5876        &$0.19\pm 0.03$  &$0.13\pm 0.02$  &$0.11\pm 0.02$  &$0.075\pm 0.014$&$0.11\pm 0.04$  &$0.08\pm 0.03$  \\
{[}\fevii ]6087&$0.12\pm 0.02$  &$0.081\pm 0.013$&$0.10\pm 0.03$  &$0.064\pm 0.020$&$0.075\pm 0.022$&$0.051\pm 0.015$\\
{[}\oi ]6300   &$0.31\pm 0.04$  &$0.20\pm 0.03$  &$0.67\pm 0.08$  &$0.41\pm 0.05$  &$0.20\pm 0.03$  &$0.13\pm 0.02$  \\
{[}\siii ]6312 &$0.05\pm 0.01$  &$0.031\pm 0.004$&$0.055\pm 0.009$&$0.033\pm 0.006$&$0.033\pm 0.008$&$0.021\pm 0.005$\\
{[}\oi ]6364   &$0.10\pm 0.01$  &$0.07\pm 0.01$  &$0.22\pm 0.03$  &$0.13\pm 0.02$  &$0.068\pm 0.015$&$0.044\pm 0.010$\\
{[}\fex ]6374  &$0.033\pm 0.009$&$0.021\pm 0.006$&  \nodata       &  \nodata       &$0.061\pm 0.016$&$0.039\pm0.010$ \\
{[}\nii ]6548  &$1.23\pm 0.17$  &$0.74\pm 0.10$  &$0.74\pm 0.11$  &$0.42\pm 0.07$  &$0.52\pm 0.08$  &$0.32\pm 0.05$  \\
H$\alpha$ 6563 &$4.63\pm 0.56$  &$2.80\pm 0.34$  &$4.93\pm 0.69$  &$2.80\pm 0.39$  &$4.53\pm 0.52$  &$2.80\pm 0.32$  \\
{[}\nii ]6583  &$3.68\pm 0.48$  &$2.21\pm 0.29$  &$2.23\pm 0.23$  &$1.26\pm 0.13$  &$1.50\pm 0.17$  &$0.92\pm 0.10$  \\
HeI6678        &$0.072\pm 0.014$&$0.04\pm 0.01$  &  \nodata       &  \nodata       &  \nodata       &  \nodata       \\
{[}\sii ]6716  &$1.18\pm 0.15$  &$0.69\pm 0.09$  &$1.09\pm 0.14$  &$0.59\pm 0.08$  &$0.26\pm 0.05$  &$0.15\pm 0.03$  \\
{[}\sii ]6731  &$1.08\pm 0.14$  &$0.63\pm 0.08$  &$0.81\pm 0.12$  &$0.44\pm 0.06$  &$0.20\pm 0.04$  &$0.12\pm 0.02$  \\
\tableline
\end{tabular}
}
\end{table*}


\begin{table*}
\setcounter{table}{4}
{\footnotesize
\caption{cont.\label{tbl-5}}
\begin{tabular}{lcccccc}
\tableline\tableline
line  &\multicolumn{2}{c}{CTS\,J13.12} & \multicolumn{2}{c}{IRAS15091-2107}
& \multicolumn{2}{c}{Mkn291}\\
      &obs.&corr.&obs.&corr.&obs.&corr.\\
(1)   & (2) & (3) & (4) & (5) & (6) & (7) \\
\tableline
{[}\nev ]3346  &  \nodata       &  \nodata       &  \nodata       &  \nodata       &  \nodata       &  \nodata      \\
{[}\nev ]3426  &$1.71\pm 0.13$  &$2.27\pm 0.17$  &$0.42\pm 0.06$  &$1.33\pm 0.18$  &$0.27\pm 0.11$  &$0.42\pm 0.17$ \\
{[}\fevii ]3588&  \nodata       &  \nodata       &  \nodata       &  \nodata       &$0.11\pm 0.04$  &$0.16\pm 0.06$ \\
{[}\oii ]3727  &$1.04\pm 0.09$  &$1.29\pm 0.11$  &$1.52\pm 0.20$  &$3.73\pm 0.49$  &$1.25\pm 0.25$  &$1.77\pm 0.35$ \\
{[}\fevii ]3760&$1.02\pm 0.09$  &$1.26\pm 0.11$  &$0.043\pm 0.009$&$0.10\pm 0.02$  &  \nodata       &  \nodata      \\
{[}\neiii ]3869&$0.88\pm 0.07$  &$1.07\pm 0.09$  &$0.61\pm 0.08$  &$1.32\pm 0.18$  &$0.38\pm 0.09$  &$0.51\pm 0.12$ \\
HeI3889        &$0.40\pm 0.05$  &$0.49\pm 0.06$  &$0.074\pm 0.016$&$0.16\pm 0.04$  &  \nodata       &  \nodata      \\
H$\zeta$3888   &$0.090\pm 0.016$&$0.11\pm 0.02$  &$0.074\pm 0.016$&$0.16\pm 0.04$  &  \nodata       &  \nodata      \\
{[}\neiii ]3968&$0.53\pm 0.05$  &$0.63\pm 0.06$  &$0.18\pm 0.06$  &$0.35\pm 0.12$  &  \nodata       &  \nodata      \\
H$\epsilon$3970&$0.11\pm 0.02$  &$0.13\pm 0.02$  &$0.036\pm 0.007$&$0.073\pm 0.014$&  \nodata       &  \nodata      \\
{[}\sii ]4072  &  \nodata       &  \nodata       &  \nodata       &  \nodata       &  \nodata       &  \nodata      \\
H$\delta$ 4101 &$0.17\pm 0.02$  &$0.20\pm 0.02$  &$0.11\pm 0.02$  &$0.20\pm 0.03$  &$0.093\pm 0.022$&$0.12\pm 0.03$ \\
H$\gamma$ 4340 &$0.31\pm 0.03$  &$0.35\pm 0.04$  &$0.34\pm 0.06$  &$0.52\pm 0.09$  &$0.27\pm 0.06$  &$0.32\pm 0.07$ \\
{[}\oiii ]4363 &$0.63\pm 0.08$  &$0.70\pm 0.09$  &$0.20\pm 0.04$  &$0.30\pm 0.06$  &$0.092\pm 0.027$&$0.11\pm 0.03$ \\
HeI4471        &$0.18\pm 0.03$  &$0.19\pm 0.03$  &$0.019\pm 0.006$&$0.026\pm 0.009$&  \nodata       &  \nodata      \\
HeII4686       &$0.50\pm 0.05$  &$0.52\pm 0.05$  &$0.21\pm 0.05$  &$0.24\pm 0.05$  &$0.08\pm 0.02$&$0.085\pm 0.020$\\
H$\beta$ 4861  &$1.00\pm 0.06$  &$1.00\pm 0.06$  &$1.00\pm 0.10$  &$1.00\pm 0.10$  &$1.00\pm 0.14$  &$1.00\pm 0.14$ \\
{[}\oiii ]4959 &$1.87\pm 0.13$  &$1.84\pm 0.13$  &$2.85\pm 0.33$  &$2.63\pm 0.31$  &$0.68\pm 0.17$  &$0.66\pm 0.17$ \\
{[}\oiii ]5007 &$5.72\pm 0.48$  &$5.55\pm 0.46$  &$8.69\pm 0.88$  &$7.70\pm 0.78$  &$2.05\pm 0.31$  &$1.95\pm 0.30$ \\
{[}\fevii ]5159&$0.36\pm 0.13$  &$0.34\pm 0.13$  &  \nodata       &  \nodata       &  \nodata       &  \nodata      \\
{[}\fevi ]5176 &  \nodata       &  \nodata       &  \nodata       &  \nodata       &  \nodata       &  \nodata      \\
{[}\Ni ]5199   &  \nodata       &  \nodata       &  \nodata       &  \nodata       &  \nodata       &  \nodata      \\
{[}\fevii ]5278&  \nodata       &  \nodata       &  \nodata       &  \nodata       &  \nodata       &  \nodata      \\
{[}\fevii ]5283&  \nodata       &  \nodata       &  \nodata       &  \nodata       &  \nodata       &  \nodata      \\
{[}\fexiv ]5307&  \nodata       &  \nodata       &  \nodata       &  \nodata       &  \nodata       &  \nodata      \\
{[}\cav ]5309  &  \nodata       &  \nodata       &  \nodata       &  \nodata       &  \nodata       &  \nodata      \\
{[}\fevii ]5721&$0.28\pm 0.04$  &$0.24\pm 0.04$  &$0.062\pm 0.020$&$0.032\pm 0.011$&  \nodata       &  \nodata      \\
{[}\nii ]5755  &  \nodata       &  \nodata       &$0.057\pm 0.009$&$0.029\pm 0.005$&  \nodata       &  \nodata      \\
HeI5876        &$0.38\pm 0.04$  &$0.32\pm 0.03$  &$0.14\pm 0.02$  &$0.07\pm 0.01$  &$0.061\pm 0.018$&$0.046\pm 0.014$\\
{[}\fevii ]6087&$0.19\pm 0.03$  &$0.16\pm 0.02$  &$0.12\pm 0.02$  &$0.054\pm 0.010$&  \nodata       &  \nodata      \\
{[}\oi ]6300   &$0.20\pm 0.02$  &$0.16\pm 0.01$  &$0.46\pm 0.06$  &$0.18\pm 0.02$  &$0.12\pm 0.02$  &$0.082\pm 0.016$\\
{[}\siii ]6312 &$0.06\pm 0.01$  &$0.05\pm 0.01$  &$0.070\pm 0.013$&$0.027\pm 0.005$&  \nodata       &  \nodata      \\
{[}\oi ]6364   &$0.067\pm 0.008$&$0.053\pm 0.007$&$0.15\pm 0.02$  &$0.06\pm 0.01$  &$0.04\pm 0.01$ &$0.027\pm 0.007$\\
{[}\fex ]6374  &$0.59\pm 0.05$  &$0.46\pm 0.04$  &  \nodata       &  \nodata       &  \nodata       &  \nodata      \\ 
{[}\nii ]6548  &$0.33\pm 0.05$  &$0.26\pm 0.04$  &$1.32\pm 0.21$  &$0.46\pm 0.08$  &$0.62\pm 0.13$  &$0.41\pm 0.08$ \\
H$\alpha$ 6563 &$3.61\pm 0.31$  &$2.80\pm 0.24$  &$8.02\pm 1.08$  &$2.79\pm 0.38$  &$4.22\pm 0.73$  &$2.80\pm 0.49$ \\
{[}\nii ]6583  &$0.99\pm 0.11$  &$0.76\pm 0.09$  &$4.34\pm 0.80$  &$1.50\pm 0.28$  &$1.86\pm 0.37$  &$1.22\pm 0.24$ \\
HeI6678        &$0.048\pm 0.011$&$0.036\pm 0.008$&$0.021\pm 0.007$&$0.007\pm 0.002$&  \nodata       &  \nodata      \\
{[}\sii ]6716  &$0.29\pm 0.02$  &$0.22\pm 0.02$  &$1.06\pm 0.15$  &$0.34\pm 0.05$  &$0.71\pm 0.11$  &$0.46\pm 0.07$ \\
{[}\sii ]6731  &$0.40\pm 0.03$  &$0.30\pm 0.02$  &$1.01\pm 0.14$  &$0.32\pm 0.05$  &$0.67\pm 0.11$  &$0.43\pm 0.07$ \\
\tableline
\end{tabular}
}
\end{table*}


\begin{table*}
\setcounter{table}{4}
{\footnotesize
\caption{cont.\label{tbl-2}}
\begin{tabular}{lcccccc}
\tableline\tableline
line  &\multicolumn{2}{c}{RXS\,J20002-5417} & \multicolumn{2}{c}{ESO\,399-IG20}
& \multicolumn{2}{c}{Mkn896}\\
      &obs.&corr.&obs.&corr.&obs.&corr.\\
(1)   & (2) & (3) & (4) & (5) & (6) & (7) \\
\tableline
{[}\nev ]3346  &$0.38\pm 0.07$  &$0.38\pm 0.07$  &  \nodata       &  \nodata       &  \nodata       &  \nodata       \\
{[}\nev ]3426  &$0.54\pm 0.08$  &$0.54\pm 0.08$  &$2.54\pm 0.21$  &$2.70\pm 0.23$  &$0.86\pm 0.29$  &$2.69\pm 0.90$  \\
{[}\fevii ]3588&  \nodata       &  \nodata       &  \nodata       &  \nodata       &$0.34\pm 0.06$  &$0.93\pm 0.16$  \\
{[}\oii ]3727  &$1.85\pm 0.29$  &$1.85\pm 0.29$  &$3.33\pm 0.26$  &$3.49\pm 0.27$  &$1.60\pm 0.33$  &$3.85\pm 0.79$  \\
{[}\fevii ]3760&$0.23\pm 0.03$  &$0.23\pm 0.03$  &$0.67\pm 0.07$  &$0.70\pm 0.07$  &$0.69\pm 0.16$  &$1.62\pm 0.36$  \\
{[}\neiii ]3869&$0.41\pm 0.11$  &$0.41\pm 0.11$  &$1.58\pm 0.14$  &$1.64\pm 0.14$  &$0.58\pm 0.27$  &$1.25\pm 0.57$  \\
HeI3889        &$0.14\pm 0.05$  &$0.14\pm 0.05$  &$0.10\pm 0.04$  &$0.11\pm 0.04$  &  \nodata       &  \nodata       \\
H$\zeta$3888   &$0.14\pm 0.05$  &$0.14\pm 0.05$  &$0.09\pm 0.03$  &$0.10\pm 0.03$  &  \nodata       &  \nodata       \\
{[}\neiii ]3968&$0.18\pm 0.04$  &$0.18\pm 0.04$  &$0.44\pm 0.08$  &$0.46\pm 0.08$  &  \nodata       &  \nodata       \\
H$\epsilon$3970&$0.18\pm 0.04$  &$0.18\pm 0.04$  &$0.12\pm 0.02$  &$0.12\pm 0.02$  &  \nodata       &  \nodata       \\
{[}\sii ]4072  &  \nodata       &  \nodata       &  \nodata       &  \nodata       &  \nodata       &  \nodata       \\
H$\delta$ 4101 &$0.25\pm 0.05$  &$0.25\pm 0.05$  &$0.22\pm 0.03$  &$0.22\pm 0.03$  &$0.16\pm 0.06$  &$0.29\pm 0.10$  \\
H$\gamma$ 4340 &$0.45\pm 0.08$  &$0.45\pm 0.08$  &$0.44\pm 0.05$  &$0.45\pm 0.05$  &$0.31\pm 0.07$  &$0.47\pm 0.11$  \\
{[}\oiii ]4363 &$0.20\pm 0.04$  &$0.20\pm 0.04$  &$1.08\pm 0.11$  &$1.10\pm 0.11$  &$0.65\pm 0.16$  &$0.98\pm 0.24$  \\
HeI4471        &  \nodata       &  \nodata       &$0.03\pm 0.02$  &$0.03\pm 0.02$&  \nodata       &  \nodata       \\
HeII4686       &$0.11\pm 0.02$  &$0.11\pm 0.02$  &$0.30\pm 0.08$  &$0.30\pm 0.08$  &$0.24\pm 0.09$  &$0.28\pm 0.11$  \\
H$\beta$ 4861  &$1.00\pm 0.11$  &$1.00\pm 0.11$  &$1.00\pm 0.04$  &$1.00\pm 0.04$  &$1.00\pm 0.17$  &$1.00\pm 0.17$  \\
{[}\oiii ]4959 &$1.40\pm 0.23$  &$1.40\pm 0.23$  &$3.92\pm 0.33$  &$3.91\pm 0.33$  &$1.61\pm 0.34$  &$1.49\pm 0.32$  \\
{[}\oiii ]5007 &$4.11\pm 0.62$  &$4.11\pm 0.62$  &$11.97\pm 0.63$ &$11.89\pm 0.62$ &$4.83\pm 0.99$  &$4.29\pm 0.88$  \\
{[}\fevii ]5159&$0.23\pm 0.07$  &$0.23\pm 0.07$  &$0.24\pm 0.06$  &$0.24\pm 0.06$  &$0.48\pm 0.14$  &$0.38\pm 0.11$  \\
{[}\fevi ]5176 &$0.25\pm 0.08$  &$0.25\pm 0.08$  &  \nodata       &  \nodata       &$0.63\pm 0.21$  &$0.48\pm 0.17$  \\
{[}\Ni ]5199   &  \nodata       &  \nodata       &  \nodata       &  \nodata       &  \nodata       &  \nodata       \\
{[}\fevii ]5278&  \nodata       &  \nodata       &  \nodata       &  \nodata       &  \nodata       &  \nodata       \\
{[}\fevii ]5283&  \nodata       &  \nodata       &  \nodata       &  \nodata       &  \nodata       &  \nodata       \\
{[}\fexiv ]5307&  \nodata       &  \nodata       &  \nodata       &  \nodata       &  \nodata       &  \nodata       \\
{[}\cav ]5309  &  \nodata       &  \nodata       &  \nodata       &  \nodata       &  \nodata       &  \nodata       \\
{[}\fevii ]5721&$0.028\pm 0.003$&$0.028\pm 0.003$&$0.20\pm 0.02$  &$0.20\pm 0.02$  &$0.44\pm 0.15$  &$0.23\pm 0.08$  \\
{[}\nii ]5755  &  \nodata       &  \nodata       &  \nodata       &  \nodata       &  \nodata       &  \nodata       \\
HeI5876        &$0.12\pm 0.03$  &$0.12\pm 0.03$  &$0.12\pm 0.02$  &$0.11\pm 0.02$  &$0.24\pm 0.06$  &$0.12\pm 0.03$  \\
{[}\fevii ]6087&$0.094\pm 0.018$&$0.094\pm 0.018$&$0.33\pm 0.04$  &$0.32\pm 0.03$  &$0.41\pm 0.18$  &$0.18\pm 0.08$  \\
{[}\oi ]6300   &$0.15\pm 0.03$  &$0.15\pm 0.03$  &$0.38\pm 0.03$  &$0.37\pm 0.03$  &$0.29\pm 0.08$  &$0.12\pm 0.03$  \\
{[}\siii ]6312 &$0.02\pm 0.01$  &$0.02\pm 0.01$  &  \nodata       &  \nodata       &  \nodata       &  \nodata       \\
{[}\oi ]6364   &$0.05\pm 0.01$  &$0.05\pm 0.01$  &$0.13\pm 0.02$  &$0.12\pm 0.02$  &$0.096\pm 0.036$&$0.037\pm 0.014$\\
{[}\fex ]6374  &$0.26\pm 0.09$  &$0.26\pm 0.09$  &$0.21\pm 0.03$  &$0.20\pm 0.03$  &$0.24\pm 0.08$  &$0.09\pm 0.03$ \\
{[}\nii ]6548  &$0.34\pm 0.06$  &$0.34\pm 0.06$  &$1.20\pm 0.15$  &$1.13\pm 0.14$  &$1.82\pm 0.61$  &$0.65\pm 0.22$ \\
H$\alpha$ 6563 &$2.64\pm 0.41$  &$2.64\pm 0.41$  &$2.96\pm 0.23$  &$2.80\pm 0.22$  &$7.86\pm 0.93$  &$2.79\pm 0.64$ \\
{[}\nii ]6583  &$1.03\pm 0.17$  &$1.03\pm 0.17$  &$3.59\pm 0.45$  &$3.40\pm 0.42$  &$5.46\pm 1.27$  &$1.92\pm 0.45$ \\
HeI6678        &  \nodata       &  \nodata       &  \nodata       &  \nodata       &$0.56\pm 0.16$  &$0.19\pm 0.05$ \\
{[}\sii ]6716  &$0.51\pm 0.08$  &$0.51\pm 0.08$  &$1.26\pm 0.10$  &$1.19\pm 0.10$  &$1.10\pm 0.23$ &$0.36\pm 0.08$ \\
{[}\sii ]6731  &$0.37\pm 0.06$  &$0.37\pm 0.06$  &$1.13\pm 0.09$  &$1.06\pm 0.09$  &$1.10\pm 0.23$ &$0.36\pm 0.08$ \\
\tableline
\end{tabular}
}
\end{table*}

\setcounter{table}{5}
\hspace*{-15mm}
\begin{deluxetable}{lccccc}
\tablewidth{0pt}
\tabletypesize{\scriptsize}
\tablecaption{Measurements of the equivalent width W$_\lambda$ of the H$\beta$ 
              emission-line profile and the individual profile components
              (in units of \AA ).}
\tablehead{
\colhead{object}     &
\colhead{W$_\lambda$(H$\beta _{tot}$)} &
\colhead{W$_\lambda$(H$\beta _{BLR}$)} &
\colhead{W$_\lambda$(H$\beta _{broad}$)} &
\colhead{W$_\lambda$(H$\beta _{interm}$)} &
\colhead{W$_\lambda$(H$\beta _{NLR}$)} \\
\colhead{(1)} &
\colhead{(2)} &
\colhead{(3)} &
\colhead{(4)} &
\colhead{(5)} &
\colhead{(6)}
}
\startdata
Mkn\,705        &$106.2\pm 5.9$&$ 96.8\pm 5.2$&$42.7\pm 1.6$&$54.1\pm 1.7$&$ 9.4\pm 0.8$\\
Mkn\,1239       &$126.8\pm 9.7$&$ 88.1\pm 5.6$&$61.2\pm 2.8$&$26.9\pm 1.2$&$38.7\pm 3.8$\\
Mkn\,734        &$ 57.2\pm 3.2$&$ 53.5\pm 2.9$&$34.7\pm 1.3$&$18.8\pm 0.7$&$ 3.7\pm 0.4$\\
NGC\,4748       &$109.4\pm 5.9$&$ 92.9\pm 5.0$&$46.1\pm 1.8$&$46.8\pm 1.8$&$16.5\pm 1.3$\\
Mkn\,783        &$ 97.1\pm 8.1$&$ 73.9\pm 5.8$&$55.6\pm 3.1$&$18.3\pm 1.0$&$23.2\pm 2.3$\\
IRAS\,13224-3809&$ 41.4\pm 2.8$&$ 25.4\pm 1.8$&$18.4\pm 0.9$&$ 7.0\pm 0.4$&$16.1\pm 1.3$\\
CTS J13.12      &$ 83.8\pm 4.4$&$ 79.2\pm 4.1$&$41.6\pm 1.5$&$37.5\pm 1.4$&$ 4.7\pm 0.4$\\
IRAS\,15091-2107&$ 93.7\pm 5.6$&$ 81.5\pm 4.5$&$49.3\pm 1.9$&$32.3\pm 1.3$&$12.1\pm 1.3$\\
Mkn\,291        &$ 28.5\pm 3.3$&$ 17.3\pm 1.5$&$12.4\pm 0.8$&$ 4.9\pm 0.2$&$11.2\pm 1.7$\\
RXS\,J20002-5417&$117.2\pm 6.5$&$109.6\pm 5.8$&$66.6\pm 2.5$&$43.0\pm 1.6$&$ 7.6\pm 0.9$\\
ESO\,399-IG20   &$101.3\pm 5.2$&$ 95.4\pm 4.9$&$86.0\pm 3.1$&$ 9.3\pm 0.4$&$ 5.9\pm 0.4$\\
Mkn\,896        &$ 54.4\pm 3.1$&$ 52.4\pm 2.9$&$25.8\pm 1.0$&$26.6\pm 1.1$&$ 2.3\pm 0.4$\\
\enddata
\end{deluxetable}

\end{document}